\documentclass{aa}

\usepackage{graphicx}
\usepackage[utf8]{inputenc}
\usepackage{amssymb}
\usepackage{amsmath}
\usepackage{graphicx}
\usepackage{xcolor}
\usepackage{enumitem}
\usepackage{url}
\usepackage{multirow}
\usepackage{comment}
\setlist{nosep}
\usepackage{tablefootnote}
\usepackage[colorlinks=true,linkcolor=blue,citecolor=blue]{hyperref}
\usepackage{ulem}

\usepackage{natbib}  
\bibpunct{(}{)}{;}{a}{}{,}

\usepackage[varg]{txfonts}

\usepackage{multicol, multirow}

\defcitealias{M19}{M19}
\defcitealias{grbpop}{S23}
\defcitealias{WP15}{W15}
\defcitealias{G16}{G16}

\begin{document}

\title{Short gamma-ray burst progenitors have short delay times}

\author{M. {Pracchia} \inst{1} \and O.~S. {Salafia} \inst{2,3}}

   \institute{STAR Institute, Université de Liège, Sart Tilman B4000 Liège, Belgium
   \and
   INAF -- Osservatorio Astronomico di Brera, via Emilio Bianchi 46, I-23807 Merate (LC), Italy
        \and
             INFN -- sezione di Milano-Bicocca, Piazza della Scienza 3, I-20126 Milano (MI), Italy
             }


   \abstract{
    Short gamma-ray bursts (SGRBs) are thought to be primarily associated with binary neutron star (BNS) mergers. The SGRB population can therefore be scrutinized for signatures of the delay time between the formation of the progenitor massive star binary and the eventual merger, which could produce an evolution of the cosmic rate density of such events whose shape departs from that of the cosmic star formation history. We therefore studied a large sample of SGRBs within a hierarchical Bayesian framework, with a particular focus on the delay time distribution (DTD) of the population. Following previous studies, we modelled the DTD either as a power-law with a minimum time delay or as a log-normal function. We considered two models for the intrinsic SGRB luminosity distribution: an empirical luminosity function with a doubly broken power-law shape and another based on a quasi-universal structured jet model. Regardless of the chosen parametrization, we found average time delays of $10\lesssim \langle \tau_\mathrm{d}\mathrm\rangle/\mathrm{Myr}\lesssim 900$ and a minimum delay time of $\tau_\mathrm{d,min}\lesssim 350\,\mathrm{Myr}$, in contrast with previous studies that found long delay times of a few gigayears. We demonstrate that the cause of the longer inferred time delays in past studies most likely resides in an incorrect treatment of selection effects.}


\label{firstpage}
\maketitle

\section{Introduction}\label{sec:intro}
Gamma-ray bursts (GRBs) are high energy gamma-ray transients of astrophysical origin. They are thought to be produced as a result of cataclysmic astrophysical events, and they are commonly classified according to their time duration and spectral hardness~\citep{Kouveliotou_1993, von_Kienlin_2020_GBM10yrs}: Short GRBs (SGRBs) are shorter in time duration and on average harder in energy spectrum compared to long GRBs (LGRBs). While a sizeable number of LGRBs have been solidly associated with extreme cases of core-collapse supernovae (CCSNe), SGRBs are believed to mainly be associated with compact binary coalescence (CBC) events, namely binary neutron star (BNS) or neutron star-black hole (NSBH) mergers \citep{Berger2014}. This connection was cemented by the joint observation and unambiguous association of GRB~170817A and the gravitational wave signal GW170817 from a BNS merger \citep{MMT_170817_obs}. 

Studies of the SGRB population generally aim to study the intrinsic luminosity distribution (also known as the `luminosity function') of those events, inferring its parameters from a sample of GRBs observed at cosmological distances by various instruments. While some studies describe the luminosity distribution as an empirically defined function~(\citealt{WP15}, hereafter \citetalias{WP15}; \citealt{G16}, hereafter \citetalias{G16}), others have tested the structured jet hypothesis by utilising more physically motivated models to describe the emitted luminosity~(\citealt{Tan_Yu_structured_jet}; \citealt{grbpop}, hereafter \citetalias{grbpop}). 

Bearing in mind the CBC origin scenario, the cosmological rate density evolution of the SGRB population depends on the time needed by the compact objects to form from massive stars and for the binary system to merge. Simple theoretical considerations for a BNS system whose orbital separation shrinks due to gravitational radiation suggest that time delays ($\tau_\mathrm{d}$) between the star formation and the merger should be distributed as a power-law $\mathrm{d}P/\mathrm{d}\tau_\mathrm{d} \propto \tau_\mathrm{d}^{-\alpha_\tau}$ with $\alpha_\tau \sim 1$~\citep{Piran_1992}. This model has been widely used in SGRB population studies (e.g.~\citealt{Paul_2018_pseudoz}, \citealt{Tan_Yu_structured_jet}, \citetalias{WP15}). For example, \citetalias{WP15}, who studied a sample of SGRBs detected by the Fermi Gamma-ray Burst Monitor (\textit{Fermi}/GBM), the Swift Burst Alert Telescope (\textit{Swift}/BAT), and the Compton Gamma-Ray Observatory Burst And Transient Source Experiment (\textit{CGRO}/BATSE), found a delay-time distribution (DTD) power-law index $\alpha_\tau \sim 1$, keeping the minimum time delay fixed at $\tau^\mathrm{min}_\mathrm{d} = 20~\mathrm{Myr}$. Some works suggest, though, that the power-law index of the DTD might be steeper, leading to shorter coalescence times. \citet{DAvanzo_2014_Swiftcomplete}, for example, selected a sample of \textit{Swift}/BAT using quality cuts that maximise the probability of a redshift measurement and found that its DTD is best represented by $\alpha_\tau \sim 1.5$ and $\tau^\mathrm{min}_\mathrm{d} = 10~\mathrm{Myr}$. More recent studies of the association between SGRB events with their host galaxies \citep{Zevin_2022} also inferred a steeper power-law DTD, with $\alpha_\tau \sim 1.5 - 2.2$, but with larger minimum time delays, $\tau^\mathrm{min}_\mathrm{d} \sim 105 - 250~\mathrm{Myr}$. An analysis of the chemical enrichment of r-process heavy elements in the Milky Way, on the other hand, under the assumption that they are mostly generated in BNS mergers, finds that fast mergers are required, with $\alpha_\tau \gtrsim 2.0$ and $\tau_\mathrm{min} \lesssim 40~\mathrm{Myr}$~\citep{Chen_2025_galacticrprocessabundancy}. 

Another widely adopted model of the SGRB DTD is a log-normal distribution $\mathrm{d}P/\mathrm{d}\tau_\mathrm{d} = \exp[-1/2(\ln\tau_\mathrm{d}-\ln\mu_\tau)^2/2\sigma_\tau^2)]/\sigma_\tau\tau_\mathrm{d}\sqrt{2\pi}$, for which~\citetalias{WP15} finds $\mu_\tau \sim 3-4~\mathrm{Gyr}$ and $\sigma_\tau \lesssim 0.2$. Similar values are reported by \citet{Luo_2022_lognormalong}, where a simulated SGRB population with a structured jet model fitting with the observed \textit{Fermi}/GBM photon flux distribution is shown to be compatible with a log-normal DTD, disfavouring power-law and Gaussian DTD models.

Some works find different evolution branches for the BNS population, which can correspond to different BNS formation channels. Studies of BNS systems in our Galaxy show that the BNS population might be composed of two distinct sub-populations: a `fast' population, with either a power-law DTD with $\alpha_\tau \sim 2$~\citep{Maoz_2025_knees} or a log-normal one with $\mu_\tau \sim 300~\mathrm{Myr}$ and $\sigma_\tau \sim 1$~\citep{Beniamini_2019_slowfast}, and a `slow' population, with $\tau_\mathrm{d}$ typically above $\sim 1~\mathrm{Gyr}$ and with $\alpha_\tau \sim 1$. In particular, \citet{Maoz_2025_knees} estimates that within Galactic BNS systems that merge within a Hubble time, the fast component is about ten to $100$ times more abundant than the slow one. \citet{Beniamini_2019_slowfast} reached a similar but slightly weaker conclusion that almost half of the Milky Way BNS population must belong to the fast-merging channel. Multi-band optical and near-infrared observations of SGRB host galaxies aimed at measuring their stellar masses and population ages \citep{Leibler_2010_duality} have shown a similar duality, where long ($\tau_\mathrm{d} \sim 3~\mathrm{Gyr}$) and short time delays ($\tau_\mathrm{d} \sim 0.2~\mathrm{Gyr}$) are respectively associated with early and late-type galaxies.

In this paper we perform a study of the SGRB population, assuming two different models of the DTD and two distinct models of the luminosity function, to test the robustness of our conclusions. The parameters are inferred within a hierarchical Bayesian framework (\citealt{M19}, hereafter \citetalias{M19}).

In Sect.~\ref{sec:time_delays} we describe our methodology and sample selection. Results are shown in Sect.~\ref{sec:results} and discussed in Sect.~\ref{sec:discussion}, where  we also demonstrate a few possible sources of biases, and we analyse their impact on inferring the parameters of our population.

\section{Methods}\label{sec:time_delays}

\subsection{Bayesian hierarchical inference method}
Our approach to characterise the SGRB population closely follows \citetalias{grbpop}, with some updates that we describe below. The parameters of the population model (formally called `hyper-parameters') constitute the elements of the hyper-parameter vector $\vec{\lambda}^\prime_\mathrm{pop}$.  Each SGRB is described by a vector of source parameters, $\vec{\lambda}_{\mathrm{src},i}$, which are estimated based on the measured data, $\vec{d}_i$ (here $i=1,...,N_\mathrm{obs}$ and $N_\mathrm{obs}$ is the number of events in the sample). 
Following Eqs.~7 and 8 in \citetalias{M19}, the posterior probability distribution function (PDF) can be written as
\begin{equation}\label{eqn:Bayes}P(\vec{\lambda}^\prime_\mathrm{pop}|\{\vec{d}_i\}) = \frac{\pi(\vec{\lambda}^\prime_\mathrm{pop})P(\{\vec{d}_i\}|\vec{\lambda}^\prime_\mathrm{pop})}{P(\{\vec{d}_i\})} = \frac{\pi(\vec{\lambda}^\prime_\mathrm{pop})}{P(\{\vec{d}_i\})}\prod^{N_\mathrm{obs}}_{i=1}  \frac{\mathcal{N}_i (\vec{d}_i|\vec{\lambda}^\prime_\mathrm{pop})}{\mathcal{D} (\vec{\lambda}^\prime_\mathrm{pop})}.
\end{equation}
Here $\pi(\vec{\lambda}_\mathrm{pop}^\prime)$ is the `hyper-prior', that is, the joint prior on the hyper-parameters, and $P(\{\vec{d}_i\})$ is the Bayesian evidence of the data (in practice, a normalisation constant). The likelihood of the data given the hyper-parameters, $P(\{\vec{d}_i\}|\vec{\lambda}^\prime_\mathrm{pop})$, is written as a product of single-source terms, each of which is expressed as a ratio of a numerator over a denominator:
\begin{equation}\label{eqn:ni_likelihood}
\mathcal{N}_i(\vec{d}_i|\vec{\lambda}^\prime_\mathrm{pop}) = \int P(\vec{d}_i|\vec{\lambda}_\mathrm{src})P_\mathrm{pop}(\vec{\lambda}_\mathrm{src}|\vec{\lambda}^\prime_\mathrm{pop})\mathrm{d}\vec{\lambda}_\mathrm{src},
\end{equation}
and
\begin{equation}\label{eqn:di_likelihood}
\mathcal{D} (\vec{\lambda}^\prime_\mathrm{pop}) = \int P_\mathrm{det}(\vec{\lambda}_\mathrm{src})P_\mathrm{pop}(\vec{\lambda}_\mathrm{src}|\vec{\lambda}^\prime_\mathrm{pop})\mathrm{d}\vec{\lambda}_\mathrm{src}.
\end{equation}
Here, $P_\mathrm{pop}(\vec{\lambda}_\mathrm{src}|\vec{\lambda}^\prime_\mathrm{pop})$ is the `population probability' that specifies the probability density of the source parameters for any choice of the hyper-parameters, and $P_\mathrm{det}(\vec{\lambda}_\mathrm{src})$ is the selection function. It represents the probability that an event with source parameters $\vec{\lambda}_\mathrm{src}$, sampled from the population, is selected and hence included in the sample. The selection function therefore contains all the information on selection effects that stem from both the detection process and from any additional selection cuts that define the sample. Further, $P(\vec{d}_i|\vec{\lambda}_\mathrm{src})$ is the likelihood of measuring the data for a single event given the source parameters. This framework has been implemented in the publicly available \texttt{grbpop} code,\footnote{https://github.com/omsharansalafia/grbpop, see \citetalias{grbpop}.} which we used as a starting point to implement different SGRB population models.

\subsection{Including the observed number of events in the inference}\label{subsec:including_Poisson}

To take into account the information on the number of events in the sample and hence include the SGRB local rate density ($R_0$) as one of the parameters of the population, we updated the definition of our posterior probability by embedding the Poissonian probability of the number of observed events. Following Eq.~11 in~\citetalias{M19}, given the total rate of astrophysical events in a population,
\begin{equation}\label{eqn:rate_astrophysical_sources}
    R(\vec{\lambda}_\mathrm{pop}) = \int_0^\infty \frac{\dot{\rho}(z,\vec{\lambda}_\mathrm{pop})}{1+z} \frac{\mathrm{d}V}{\mathrm{d}z} \mathrm{d}z,
\end{equation}
with $\vec{\lambda}_\mathrm{pop} \equiv (\vec{\lambda}^\prime_\mathrm{pop}, R_0)$, the number of expected detections for a sample of a given detector is
\begin{equation}\label{eqn:ndet_poisson}
    N_\mathrm{det}(\vec{\lambda}_\mathrm{pop}) = \eta_\mathrm{DC} \cdot T \cdot R(\vec{\lambda}_\mathrm{pop}) \cdot \mathcal{D}(\vec{\lambda}^\prime_\mathrm{pop}),
\end{equation}
where $\eta_\mathrm{DC}$ is the product of the detector duty cycle times the average fraction of the sky accessible to it,\footnote{The term $\eta_\mathrm{DC}$ in general expresses the fraction of events that pass any pre-selection that does not depend on the properties of the source and is not accounted for in the $P_\mathrm{det}$ selection effects model.} $R(\vec{\lambda}_\mathrm{pop})$ is the total cosmic rate of events, $T$ is the duration of the observation period, and $\mathcal{D}(\vec{\lambda}^\prime_\mathrm{pop})$ is the term defined in Eq.~\ref{eqn:di_likelihood}, which can be shown to be equivalent to the fraction of events in the population that pass the detection and selection cuts. The probability of observing $N_\mathrm{obs}$ events given $N_\mathrm{det}(\vec{\lambda}_\mathrm{pop})$ is then a Poissonian probability, and the posterior PDF becomes
\begin{equation}\label{eqn:posterior_m19_poisson}
    P(\vec{\lambda}_\mathrm{pop}|\{\vec{d}_i\}) = \frac{\pi(\vec{\lambda}^\prime_\mathrm{pop})}{P(\{\vec{d}_i\})} \prod^{N_\mathrm{obs}}_{i=1}  \left[ \frac{\mathcal{N}_i (\vec{d}_i|\vec{\lambda}^\prime_\mathrm{pop})}{\mathcal{D} (\vec{\lambda}^\prime_\mathrm{pop})} \right] e^{-N_\mathrm{det}} (N_\mathrm{det})^{N_\mathrm{obs}},
\end{equation}
where the $\frac{1}{N_\mathrm{obs}!}$ term is omitted due to the distinguishability of the events in the sample \citepalias{M19}. We defined the isotropic equivalent of the peak luminosity, $L(p_{[E_0, E_1]}, E_\mathrm{p}, z)$, following Eq.~17 in \citetalias{grbpop} and modelling the SGRB photon spectrum as a power-law with an exponential cut-off~\cite{Ghirlanda2004_GRBspectrum}, with the low-energy photon index set to the median value of $\alpha$ based on the spectral analysis of the \textit{Fermi}/GBM SGRB catalogue~\citepalias{grbpop}.

\subsection{Population models}\label{subsec:models}

\subsubsection{Rate density evolution}\label{subsubsec:rate_density_evolution}
Guided by the expectation that short GRBs are generated mainly in compact binary mergers, such as BNS or NSBH systems, the redshift probability distribution was modelled considering a fixed cosmic star formation history (CSFH) and by convolving it with the distribution of time delays ($\tau_\mathrm{d}$) between the binary system formation and the eventual merger. We considered two DTD models: a power law with a minimum time delay cut-off, namely
\begin{equation}\label{eqn:dtd_pow}
     P( \tau_\mathrm{d} |\vec{\lambda}^\prime_\mathrm{pop}) \propto
\begin{cases} 
        \displaystyle 0~, & \tau_\mathrm{d} <\tau_\mathrm{d}^\mathrm{min};  \\
        \displaystyle \tau_\mathrm{d}^{-\alpha_\tau}, & \tau_\mathrm{d} \geq \tau_\mathrm{d}^\mathrm{min}, 
\end{cases}
\end{equation}
where the free parameters are $\tau^\mathrm{min}_\mathrm{d}$ and the index $\alpha_\tau$, and a log-normal DTD, that is
\begin{equation}\label{eqn:dtd_log}
    P( \tau_\mathrm{d} |\vec{\lambda}^\prime_\mathrm{pop} ) = \exp \left[ - \frac{1}{2} \left( \frac{\ln \tau_\mathrm{d} - \ln \mu_\tau}{\sigma_\tau} \right)^2 \right] (\tau_\mathrm{d} \sqrt{2 \pi \sigma_\tau^2})^{-1},
\end{equation}
where the parameters to constrain are the median value $\mu_\tau$ and the dispersion $\sigma_\tau$. The rate density distribution is therefore
\begin{equation}\label{eqn:dtd*sfh_integral}
   \displaystyle \dot{\rho}(z,\vec{\lambda}^\prime_\mathrm{pop}) \propto \int_z^\infty \psi(z) P \left( t_\mathrm{LB}(z^\prime)-t_\mathrm{LB}(z) | \vec{\lambda}^\prime_\mathrm{pop} \right) \frac{\mathrm{d}t}{\mathrm{d}z^\prime} \mathrm{d}z^\prime,
\end{equation}
where $\psi(z)$ is the CSFH and $t_\mathrm{LB}(z)$ is the cosmological lookback time at redshift $z$. We chose to adopt the CSFH model from~\cite{MadauFragos2017}, 
\begin{equation}\label{eqn:md14_sfh}
    \psi(z) \propto \frac{(1+z)^{a_\psi}}{1+ \left( \frac{1+z}{1+z_\psi} \right)^{b_\psi}},
\end{equation}
with $a_\psi = 2.6$, $b_\psi=6.2$, and $z_\psi = 2.2$. The redshift probability distribution of SGRBs is then computed as 
\begin{equation}\label{eqn:p_z|lambdapop}
   \displaystyle P(z|\vec{\lambda}^\prime_\mathrm{pop}) \propto \frac{\dot{\rho}(z, \vec{\lambda}^\prime_\mathrm{pop})}{1+z} \frac{\mathrm{d}V}{\mathrm{d}z},
\end{equation}
where $\mathrm{d}V/\mathrm{d}z$ is the differential comoving volume at redshift $z$.

\subsubsection{Luminosity function models: ELF and QUSJ}
To describe the luminosity function (i.e. the luminosity PDF) of our population, we chose two different models. The first one, which we refer to as the `empirical luminosity function' model (ELF hereafter), is based on a doubly broken power-law luminosity function, and it extends the \citetalias{WP15} luminosity function model to lower luminosities. 
Following\footnote{We note that the formulation in  \citet{O3bGRB}, which is inspired by \citetalias{WP15}, is in terms of the probability density of the logarithm of the luminosity, $dP/d\ln L$, while in our formalism the quantity $P(L\,|\,\vec{\lambda}^\prime_\mathrm{pop})$ is a probability density of the luminosity, $dP/dL$. For that reason, the power-law slopes in our formulation are steeper by one with respect to their definition.} \citet{O3bGRB}, we express the luminosity function in this model as
\begin{equation}\label{eqn:lumfun_bpl_breaks}
    \displaystyle P(L|\vec{\lambda}^\prime_\mathrm{pop}) \propto
    \begin{cases} 
        0 & L<L_0;\\
        \displaystyle \left(\frac{L_{**}}{L_*} \right)^{-\alpha_\mathrm{BPL}-1} \left( \frac{L}{L_{**}} \right)^{-\gamma_\mathrm{BPL}-1}  , & L_0 \leq L \leq L_{**}; \\
        \displaystyle \left( \frac{L}{L_*} \right)^{-\alpha_\mathrm{BPL}-1} , & L_{**} < L \leq L_*; \\
        \displaystyle \left( \frac{L}{L_*} \right)^{-\beta_\mathrm{BPL}-1}, & L > L_*.
    \end{cases}
\end{equation}

Here $L_0$, $L_{**}$, and $L_*$ are the minimum luminosity, low-luminosity break, and high-luminosity break, respectively. The power-law indices of the three branches are expressed through the $\gamma_\mathrm{BPL}$, $\alpha_\mathrm{BPL}$, and $\beta_\mathrm{BPL}$ parameters to facilitate comparison with \citet{O3bGRB}. As a way to incorporate some information from GRB~170817A in this model, we chose our prior on $L_0$ to ensure that $L_0<L_\mathrm{17A}$, where $L_\mathrm{17A}=1.2_{-0.4}^{+0.5}\times 10^{47}\,\mathrm{erg\,s^{-1}}$ is the peak luminosity of GRB~170817A \citepalias{grbpop}. 

To fully specify the population properties, the PDF of the peak photon energy $E_\mathrm{p}$ of the GRB spectral energy distribution across the population must also be specified. Following \citetalias{grbpop}, we modelled this as a log-normal distribution with a median value that depends on $L$: 
\begin{equation}\label{eqn:ep_dispersion_bpl}
    P(E_\mathrm{p}|L, \vec{\lambda}^\prime_\mathrm{pop}) = \frac{\exp \left[ - \frac{1}{2} \left( \frac{\ln(E_\mathrm{p}) - \ln(\tilde{E}_\mathrm{p}(L) )}{\sigma_\mathrm{c}} \right)^2 \right]}{E_\mathrm{p} \sqrt{2 \pi \sigma_\mathrm{c}^2}},
\end{equation}
where $\tilde{E}_\mathrm{p} (L) = E_\mathrm{p}^\star \left( L/L_\mathrm{*} \right)^y$ allows for a Yonetoku-like correlation~\citep{yonetoku2004} when $y \neq 0$. The source parameters $L$ and $E_\mathrm{p}$ are therefore parametrized with a common probability density: $P(L, E_\mathrm{p} | \vec{\lambda}^\prime_\mathrm{pop}) = P(E_\mathrm{p}|L, \vec{\lambda}^\prime_\mathrm{pop}) P(L|\vec{\lambda}^\prime_\mathrm{pop})$.

The second model we considered is the quasi-universal structured jet population model (QUSJ hereafter) from \citetalias{grbpop}, which also features a correlated joint distribution of $L$ and $E_\mathrm{p}$, induced by the jet structure. The latter is defined by two dimensionless functions of the ratio of the viewing angle, $\theta_\mathrm{v}$, to the `jet core' angle $\theta_\mathrm{c}$, written as $\ell(\theta_\mathrm{v}/\theta_\mathrm{c})$ and $\eta(\theta_\mathrm{v}/\theta_\mathrm{c})$, that describe how the luminosity and $E_\mathrm{p}$ depend on the jet viewing angle (see \citetalias{grbpop} for more details). In the analysis based on the QUSJ model, we include the information on the luminosity, peak photon energy, and viewing angle of GRB~170817A in the form of priors on the population parameters, following \citetalias{grbpop}.

In both cases the redshift probability distribution, $P(z|\vec{\lambda}^\prime_\mathrm{pop})$, is assumed to be independent of $L$ and $E_\mathrm{p}$.
 The distribution of source parameters across the population is therefore
\begin{equation}\label{eqn:ppop_bpl}
    P_\mathrm{pop}(\vec{\lambda}_\mathrm{src} | \vec{\lambda}^\prime_\mathrm{pop}) = P(L, E_\mathrm{p} | \vec{\lambda}^\prime_\mathrm{pop}) P(z | \vec{\lambda}^\prime_\mathrm{pop})~.
\end{equation}

\subsubsection{W15 population model}

To gain insight on the difference between our results and those of previous studies, we set out to reproduce the \citetalias{WP15} constraints using hierarchical Bayesian inference but with the same sample and selection effects modelling used in that work. For this purpose, the luminosity probability distribution $P(L|\vec{\lambda}^\prime_\mathrm{pop})$ we employed is the same as in Eq.~\ref{eqn:lumfun_bpl_breaks}, but we fixed $L_0 = L_{**} = 5 \times 10^{49}~\mathrm{erg}~\mathrm{s}^{-1}$. The parameters to be constrained for $P(L|\vec{\lambda}^\prime_\mathrm{pop})$ therefore are only $\alpha_\mathrm{BPL}$, $\beta_\mathrm{BPL}$, and $L_*$.

The redshift probability distribution, $P(z|\vec{\lambda}^\prime_\mathrm{pop})$, was modelled as in Sect.~\ref{subsubsec:rate_density_evolution}. The only difference is for the power-law DTD model, where the minimum time delay was fixed to $\tau^\mathrm{min}_\mathrm{d} = 20~\mathrm{Myr}$. The SGRB photon spectrum $\mathrm{d}\dot{N}/\mathrm{d}E(E,E_\mathrm{p})$ was modelled as a Band function~\cite{Band}, whose parameters were fixed to $E_\mathrm{p}  = 800~\mathrm{keV}$, $\alpha_\mathrm{Band} = -0.5$, and $\beta_\mathrm{Band} = -2.25$, as in the original paper \citep[motivated by the analysis of][]{Nava2011}.

Since the spectral energy peak in this model is fixed, the source parameters are only the GRB luminosity and redshift, that is, $\vec{\lambda}_\mathrm{src} = (L, z)$. In Appendix \ref{app:WP15_remastered} we give technical details on how this difference is handled in the inference.

\subsection{Samples and detection efficiencies}~\label{subsec:samples}
When building our reference SGRB samples, we needed to carefully model the selection effects that moulded them. We therefore considered two samples for the analysis, built in the same way as in the `flux-limited sample analysis' of \citetalias{grbpop}.

The likelihood numerators, $\mathcal{N}_i(\vec{d}_i|\vec{\lambda}^\prime_\mathrm{pop})$, for each event in the samples and denominators, $\mathcal{D} (\vec{\lambda}^\prime_\mathrm{pop})$, were then computed in the same way as in \citetalias{grbpop} for both the ELF and the QUSJ case since the only term that changes between the two models is $P_\mathrm{pop}(\vec{\lambda}_\mathrm{src} | \vec{\lambda}^\prime_\mathrm{pop})$.

\subsubsection{Observer-frame sample}
The first sample consists of short GRBs detected by \textit{Fermi}/GBM with publicly available spectral analysis results. We selected events with $T_{90} < 2~\mathrm{s}$ from the GBM's first 10 years of observation, namely from the 12 July 2008 to 11 July 2018. We applied a completeness cut on the $64~\mathrm{ms}$ peak photon flux in the $[50-300]~\mathrm{keV}$ energy band, selecting only the events with $p_{[50, 300]} > p_\mathrm{lim, GBM} = 3.50~\mathrm{cm}^{-2}~\mathrm{s}^{-1}$, which is the value above which we can consider our sample as complete in flux \citepalias{grbpop}. Additional cuts on the observed spectral peak energy ($E_\mathrm{p,obs}$) were applied, considering only events with best-fit values $ 50~\mathrm{keV} \leq E_\mathrm{p,obs} \leq 10~\mathrm{MeV}$, which is the spectral range where the effective area of the \textit{Fermi}/GBM detectors is optimal. After all cuts, the sample contained 210 short GRB events.
Hereon, we refer to this as the `observer-frame sample'.

The detection efficiency for this sample was modelled as
\begin{equation}
\begin{split}
    P_{\mathrm{det}}(L, E_\mathrm{p},z) = \Theta\left(p_{[50, 300]}(L,E_\mathrm{p},z) - p_\mathrm{lim,GBM}\right)\times\\
    \Theta\left(E_\mathrm{p}/(1+z)-50\,\mathrm{keV}\right)\Theta\left(10\,\mathrm{MeV}-E_\mathrm{p}/(1+z)\right),
\end{split}
\end{equation}
where $\Theta(x)$ is the Heaviside step function.

Given the number of events contained in the observer-frame sample, we considered its event detection rate to estimate the astrophysical local rate density ($R_0$) of the SGRB population. The $N_\mathrm{det}$ for the Poissonian term in Eq.~\ref{eqn:posterior_m19_poisson} was therefore computed using the $\mathcal{D} (\vec{\lambda}^\prime_\mathrm{pop})$ for this sample, an observation period of $T = 10~\mathrm{yr}$, and a multiplicative factor of $\eta_\mathrm{DC} = 0.59$ accounting for both the accessible field of view and the duty cycle of \textit{Fermi}/GBM~\citep{Burns_16_dutycycle}.

\subsubsection{Rest-frame sample}
The second sample consist of short GRBs detected by \textit{Fermi}/GBM and \textit{Swift}/BAT that pass a set of stringent cuts that maximise the redshift completeness without distorting the redshift distribution. In practice, this is the sub-sample of the S-BAT4ext catalogue \citep{SBAT4ext} of events that were also detected by \textit{Fermi}/GBM. The selection includes a \textit{Swift}/BAT completeness flux threshold in the $[15-150]~\mathrm{keV}$ energy band, $p_{[15,150]} > p_\mathrm{lim, BAT} = 3.50~\mathrm{cm}^{-2}~\mathrm{s}^{-1}$, in addition to the same \textit{Fermi}/GBM completeness cuts as in the observer-frame sample. The sample consists of 18 short GRBs events, 16 of which have a measured redshift. For those events we used the results of the spectral analysis performed in \citetalias{grbpop}, which yields, for each event $i$, a posterior on the source parameters $P(L, E_\mathrm{p}, z\, | \,d_i)$, given a prior $\pi(L, E_\mathrm{p}, z) \propto (1+z)^{-1} L^{-1}$. We refer to this sample as the `rest-frame sample'. We did not use the number of events in this sample to inform our estimate of $R_0$ because of the difficulty in correctly modelling the duty cycle factor $\eta_\mathrm{DC}$ incorporating all the cuts that define the sample.

Because we required the detection by \textit{Fermi}/GBM, the rest-frame sample detection efficiency includes the same cuts as the observer-frame sample. In addition, it features a factor
\begin{equation}\label{eqn:det_hard_threshold}
    \Theta\left(p_{[15, 150]}(L,E_\mathrm{p},z) - p_\mathrm{lim,BAT}\right)
\end{equation}
to account for the \textit{Swift}/BAT completeness cut. All other S-BAT4 cuts are independent on the source properties and hence do not enter the definition of this efficiency.

We note that GRB~170817A is not included in either of the two samples because its peak flux is below the \textit{Fermi}/GBM completeness threshold. In the case of the quasi-universal jet population model, our results still include information on this event in the form of a prior on the hyper-parameters, which we built following the procedure used by \citetalias{grbpop} in their flux-limited sample analysis.

\subsubsection{W15 samples}
The observer-frame samples considered for the~\citetalias{WP15} population analysis consist of short GRBs observed by \textit{Fermi}/GBM, \textit{CGRO}/BATSE, and \textit{Swift}/BAT. For the three detectors, the detection efficiency was modelled as a hard threshold on the $64~\mathrm{ms}$ peak photon flux in their respective most sensitive detector energy band, as in Eq.~\ref{eqn:det_hard_threshold}. The photon flux thresholds considered for \textit{CGRO}/BATSE and \textit{Fermi}/GBM in the $[E_0,~E_1] = [50,~300]~\mathrm{keV}$ detector energy band for this analysis were, respectively, $p_\mathrm{lim, BATSE} = 1.5~\mathrm{cm}^{-2}~\mathrm{s}^{-1}$ and $p_\mathrm{lim, GBM} = 2.37~\mathrm{cm}^{-2}~\mathrm{s}^{-1}$, while for \textit{Swift}/BAT it was $p_\mathrm{lim, BAT} = 2.5~\mathrm{cm}^{-2}~\mathrm{s}^{-1}$ in the $[E^\mathrm{d}_0,~E^\mathrm{d}_1] = [15,~150]~\mathrm{keV}$ band.

The BATSE and \textit{Fermi} sample contain GRBs with a measured $p_{[E_0, E_1]}$ above the respective detector threshold and $T_{90} < 2~\mathrm{s}$. While for BATSE these selection effects have been applied to the entirety of its catalogue, for a total of 414 events, the \textit{Fermi}/GBM events were considered up to the 10 April 2013, as in \citetalias{WP15}, yielding 146 bursts.

The rest-frame sample was built following \citetalias{WP15}. The sample contains events detected by \textit{Swift}/BAT with a measured redshift and $L$, with $T_{90} < 2~\mathrm{s}$, up to GRB~131004A. A cut on the probability of having a non-collapsar origin higher than $60\%$, estimated following~\cite{bromberg13}, was then applied, selecting a sub-sample of 12 events.

Since the spectral peak energy in this analysis is fixed to a punctual value in the source frame, the computation of the likelihood was a bit different from that in the ELF and QUSJ cases. Full details on the likelihood evaluation and the obtained results are described in Appendix~\ref{app:WP15_remastered}. In the following, we refer to the results of this analysis as \citetalias{WP15}*.

\subsection{Priors and posterior evaluation}\label{subsec:choice_of_priors}

\begin{table*}
    \caption{Full set of population parameters and the associated priors.}
    \centering
\begin{tabular}{ c l l }
  Parameter & Prior & Description \\ 
 \hline
 \multicolumn{3}{c}{Empirical luminosity function model} \\
 \hline
 $\alpha_\mathrm{BPL}$ & Uniform, $\alpha_\mathrm{BPL} \in [-1,~5]$ & \\
 $\beta_\mathrm{BPL}$ & Uniform, $\beta_\mathrm{BPL} \in [0,~5]$ & Slopes of the luminosity PDF in the $\mathrm{d}P/\mathrm{d}\ln L$ representation \\
 $\gamma_\mathrm{BPL}$ & Uniform, $\gamma_\mathrm{BPL} \in [-5,~5]$ &\\
 $L_0$ & Uniform-in-log, $L_0 \in [10^{44},~10^{47}]~\mathrm{erg}~\mathrm{s}^{-1}$ & Minimum luminosity \\ 
 $L_{**}$ & Uniform-in-log, $L_{**} \in [L_0,~10^{51}~\mathrm{erg}~\mathrm{s}^{-1}]$ & Low-luminosity break \\ 
 $L_*$ & Uniform-in-log, $L_* \in [10^{51},~10^{56}]~\mathrm{erg}~\mathrm{s}^{-1}$ & High-luminosity break \\  
 $E_\mathrm{p}^\star$ & Uniform-in-log, $E_\mathrm{p}^\star \in [10^{2},~10^{5}]~\mathrm{keV}$ & Central $E_\mathrm{p}$ at $L = L_{*}$ in $L-E_\mathrm{p}$ correlation\\ 
 \hline
 \multicolumn{3}{c}{Quasi-universal structured jet model} \\
 \hline
 $\theta_\mathrm{c}$ & Isotropic, $\theta_\mathrm{c} \in [0.01, \pi/2]~\mathrm{rad}$ & Core half-opening angle\\
 $\theta_\mathrm{w}$ & Isotropic, $\theta_\mathrm{w} \in [\theta_\mathrm{c}, \pi/2~\mathrm{rad}]$ & Structure `break' angle\\
 $\alpha_L$ & Uniform, $\alpha_\mathrm{L} \in [0,~6]$ & Slope of $\mathcal{\ell}$ at intermediate viewing angles $\theta_\mathrm{c} \leq \theta_\mathrm{v} < \theta_\mathrm{w}$ \\
 $\beta_L$ & Uniform, $\beta_L \in [-3,~6]$ & Slope of $\mathcal{\ell}$ at large viewing angles $\theta_\mathrm{v} \geq \theta_\mathrm{w}$ \\
 $L^\star_\mathrm{c}$ & Uniform-in-log, $L^\star_\mathrm{c} \in [3 \times 10^{51},~10^{55}]~\mathrm{erg}~\mathrm{s}^{-1}$ & Typical luminosity for on-axis observers $(\theta_\mathrm{v} = 0)$ \\
 $A$ & Uniform, $A \in [1.5,~5]$ & Slope of $L_\mathrm{c}$ distribution \\
 $E_\mathrm{p,c}^\star$ & Uniform-in-log, $E_\mathrm{p, c}^\star \in [10^{2},~10^{5}]~\mathrm{keV}$ & Typical $E_\mathrm{p}$ for on-axis observers $(\theta_\mathrm{v} = 0)$ \\
 $\alpha_{E_\mathrm{p}}$ & Uniform, $\alpha_{E_\mathrm{p}} \in [0,~6]$ & Slope of $\mathcal{\eta}$ at intermediate viewing angles $\theta_\mathrm{c} \leq \theta_\mathrm{v} < \theta_\mathrm{w}$ \\
 $\beta_{E_\mathrm{p}}$ & Uniform, $\beta_{E_\mathrm{p}} \in [-3,~6]$ & Slope of $\mathcal{\eta}$ at large viewing angles $\theta_\mathrm{v} \geq \theta_\mathrm{w}$ \\  
 \hline
 \multicolumn{3}{c}{Common parameters} \\
 \hline
 $y$ & Uniform, $y \in [-3,~3]$ & $L-E_\mathrm{p}$ correlation slope \\
 $\sigma_\mathrm{c}$ & Uniform-in-log, $\sigma_\mathrm{c} \in [0.3,~3]$ & $E_\mathrm{p}$ dispersion in $L-E_\mathrm{p}$ correlation \\  
 $\tau^\mathrm{min}_\mathrm{d}$ & Uniform-in-log, $\tau^\mathrm{min}_\mathrm{d} \in [0.005,~3]~\mathrm{Gyr}$ & Minimum delay between progenitor formation and SGRB\\  
 $\alpha_\tau$ & Uniform, $\alpha_t \in [0,~5]$ & Slope of power-law DTD \\  
 $\mu_\tau$ & Uniform-in-log, $\mu_t \in [0.01,~5]~\mathrm{Gyr}$ & Median of the log-normal DTD \\  
 $\sigma_\tau$ & Uniform-in-log, $\sigma_t \in [0.01,~5]$ & Standard deviation for the log-normal DTD \\ 
 $R_0$ & Uniform-in-log, $R_0 \in [1,~10^4]~\mathrm{Gpc}^{-3}~\mathrm{yr}^{-1}$ & Short GRB local rate density \\
 \hline
\end{tabular}
    \label{tab:priors}
\end{table*}

We chose broad priors on most parameters, considering either a uniform or a uniform-in-log distribution within a defined range for each parameter, with the following exceptions: (i) For $\theta_\mathrm{c}$ and $\theta_\mathrm{w}$, we considered an isotropic prior (i.e. uniform in the subtended solid angle) $\pi(\theta_\mathrm{c/w}) \propto \sin(\theta_\mathrm{c/w})$. (ii) For the minimum luminosity of the ELF model, we chose the upper bound of the prior based on the luminosity of GRB~170817A, as stated previously. (iii) For the QUSJ model, we followed the approach of \citetalias{grbpop} to condition the prior on the observed properties of GRB~170817A (see their section 2.5.3). The bounds that we report in Table \ref{tab:priors} refer to the priors before that conditioning. The population parameters were considered to be independent from each other, with the exceptions of $(L_0, L_{**}, L_*)$ for the ELF model and $(\theta_\mathrm{c}, \theta_\mathrm{w})$ for the QUSJ model. In fact, for the triplet of luminosity breaks, we imposed $L_0 < L_{**} < 10^{51}~\mathrm{erg}~\mathrm{s}^{-1} < L_*$, while for the break angles in the quasi-universal jet structure, we set $\theta_\mathrm{c} < \theta_\mathrm{w}$. The choice to put a boundary between $L_{**}$ and $L_*$ at $L = 10^{51}~\mathrm{erg}~\mathrm{s}^{-1}$ was motivated by wanting to avoid that those parameters have local peaks in the posteriors in the same parameter region. Common parameters in the two population models share common priors. The full set of priors for each parameter with their bounds are shown in Table~\ref{tab:priors}.

The posterior PDF was estimated through dynamical nested sampling using the open source python package \texttt{dynesty}~\citep{dynamicalnestedsampling, dynesty}. The number of initial live points was set to 100 times the dimensionality of the posterior PDF, for a total of $1200$ and $1400$ initial live points, respectively, for the ELF and the QUSJ models. As stopping criteria, we chose dlogz=$10^{-5}$ and 30000 effective samples. Batches of live points were periodically and automatically added to reach those criteria, with each batch containing one fifth of the initial live points used for the sampling. The obtained posterior PDFs for all the cases considered are portrayed as corner plots in Appendix~\ref{app:corners}, with contours representing the $90\%$ and $50\%$ confidence levels, while the estimated median values of the parameters with their $90\%$ credible intervals are shown in Table 2.

\section{Results}\label{sec:results}

A summary of the constraints obtained on the population parameters for the ELF and QUSJ models, with both the power-law and log-normal DTDs, is given in Table \ref{tab:inferred_parameters}. The results for the \citetalias{WP15}* analysis are reported in Appendix \ref{app:WP15_remastered}, where we demonstrate a good agreement with the results in the original paper.

\subsection{Time delay and redshift distribution}\label{subsec:time_delays}

\begin{figure*}
\includegraphics[width=0.49\textwidth]{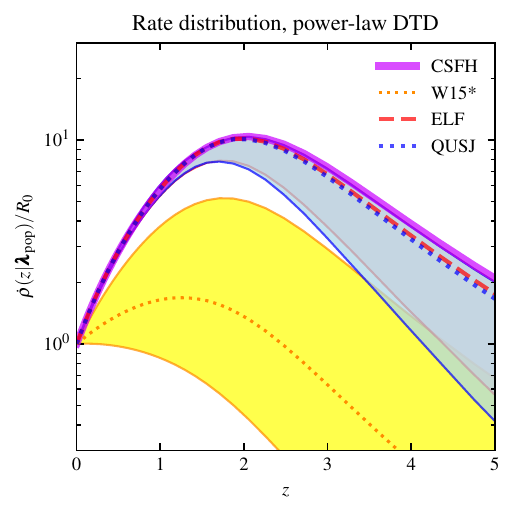}%
\includegraphics[width=0.5\textwidth]{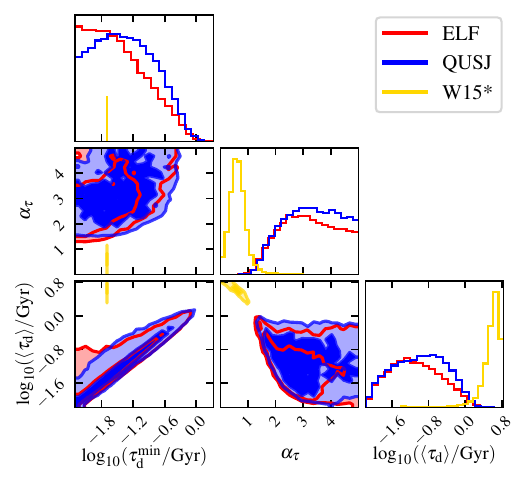}\\
\includegraphics[width=0.49\textwidth]{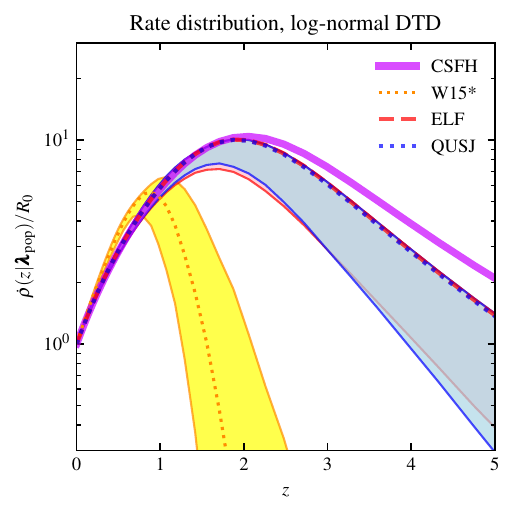}%
\includegraphics[width=0.5\textwidth]{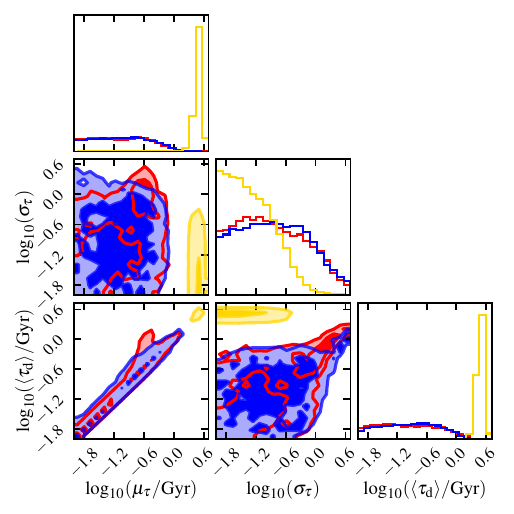}

\caption{Rate density distributions as functions of redshift (left panels) and corner plots of the posterior PDFs for the DTD parameters and $\langle \tau_\mathrm{d} \rangle$ (right panels). Top and bottom panels show the results obtained with a power-law and a log-normal DTD, respectively. Rate density distributions are normalised to $1$ at $z=0$, and the CSFH from \citet{MadauFragos2017} is shown in purple for comparison. Red and blue curves represent the distributions obtained considering, respectively, the ELF and the QUSJ models. \citetalias{WP15}* results are displayed in yellow.}
\label{fig:rate_densities}
 
\end{figure*}

We illustrate the constraints obtained with our analyses on the shape of the rate density evolution with redshift in Fig.~\ref{fig:rate_densities} (left-hand panels), where the $\dot\rho(z)/R_0$ curves (i.e.\ the SGRB rate density normalised to its value at $z=0$) are compared with the normalized CSFH. The results are quite consistent between the ELF and QUSJ models, where the curves tend to closely follow the CSFH, while the \citetalias{WP15}* analysis yields a very different constraint. The DTD parameter constraints (right-hand panels) show that this is the result of much shorter time delays in our models with respect to those obtained in the \citetalias{WP15}* analysis. To make this more evident and to enable a comparison between different DTD models, we also show in the figure, in addition to the posterior probability on the DTD parameters, the distribution of the average time delay: 
\begin{equation}\label{eqn:expected_time_delays}
    \langle \tau_\mathrm{d}\rangle(\vec{\lambda}^\prime_\mathrm{pop}) = \int \tau_\mathrm{d} P(\tau_\mathrm{d}|\vec{\lambda}^\prime_\mathrm{pop})\, \mathrm{d}\tau_\mathrm{d}.
\end{equation}

In the power-law DTD scenario, both luminosity function models show a steep power-law index $\alpha_\tau > 1.5$, while the minimum time delays are below $\tau_\mathrm{d} < 370~\mathrm{Myr}$ (90\% credible limits). The average time delays are generally below $\sim 1~\mathrm{Gyr}$, with a median and 90\% credible range of $\langle\tau_\mathrm{d}\rangle = 77_{-65}^{+619}~\mathrm{Myr}$ (ELF) or $\langle\tau_\mathrm{d}\rangle = 105_{-93}^{+665}~\mathrm{Myr}$ (QUSJ). The time delays associated with the \citetalias{WP15}* DTD parameters, on the other hand, were found to be significantly larger, yielding $\langle\tau_\mathrm{d}\rangle = 4.2_{-2.8}^{+2.0}~\mathrm{Gyr}$.

Similar results were obtained in the log-normal DTD case, where the median value of the log-normal distribution is $\mu_\tau < 700~\mathrm{Myr}$, while $\sigma_\mathrm{c}$ is only loosely constrained to be somewhat below the upper end of its prior range, which is equivalent to 1.5 dex. For the ELF and QUSJ models, the average time delays are respectively $\langle\tau_\mathrm{d}\rangle \sim 92_{-79}^{+877}~\mathrm{Myr}$ and $\langle\tau_\mathrm{d}\rangle \sim 116_{-102}^{+761}~\mathrm{Myr}$, while the \citetalias{WP15}* analysis yields $\langle\tau_\mathrm{d}\rangle = 3.1_{-0.8}^{+0.7}~\mathrm{Gyr}$.

\subsection{Luminosity function}

\begin{table*}
    \centering
        \caption{Constraints on population parameters.}
        \small\centering
\begin{tabular}{ l c c | l c c c }
 \multicolumn{3}{c}{Empirical luminosity function} &  \multicolumn{4}{c}{Quasi-universal structured jet}\\
 \hline
 Parameter & Power-law DTD & Log-normal DTD & Parameter & Power-law DTD & Log-normal DTD & S23 \\ 
 \hline
 $\alpha_\mathrm{BPL}$ & ${0.90}_{-0.36}^{+0.38}$ & ${0.81}_{-0.32}^{+0.38}$ &        $\alpha_L$ & ${4.93}_{-1.29}^{+0.96}$ & ${5.01}_{-1.20}^{+0.90}$ & $4.9^{+1.0}_{-1.7}$ \\  
 $\beta_\mathrm{BPL}$ & ${2.18}_{-0.99}^{+2.14}$ & ${2.40}_{-1.06}^{+2.24}$ &     $\beta_L$ & ${1.95}_{-4.49}^{+3.64}$ & ${1.57}_{-4.14}^{+4.00}$ & $1.9^{+3.7}_{-4.4}$ \\  
 $\gamma_\mathrm{BPL}$ & ${-2.37}_{-2.32}^{+2.22}$  & ${-1.90}_{-2.64}^{+2.97}$ &       $A$ & ${2.78}_{-0.40}^{+0.67}$  & ${2.80}_{-0.40}^{+0.65}$ & $2.9^{+0.7}_{-0.4}$ \\  
 $\log_{10}(L_* / \mathrm{erg\,s^{-1}})$ & ${53.08}_{-0.86}^{+2.26}$ & ${52.99}_{-1.03}^{+1.48}$ &        $\log_{10}(L^\star_\mathrm{c} / \mathrm{erg\,s^{-1}})$ & ${51.74}_{-0.23}^{+0.46}$ & ${51.75}_{-0.25}^{+0.43}$ & $51.70^{+0.38}_{-0.10}$ \\ 
 $\log_{10}(L_{**} / \mathrm{erg\,s^{-1}})$ & ${49.07}_{-2.15}^{+0.86}$ & ${47.67}_{-2.16}^{+1.98}$ &      $\log_{10}(E_\mathrm{p,c}^\star /\mathrm{keV})$ & ${3.50}_{-0.28}^{+0.26}$ & ${3.52}_{-0.27}^{+0.27}$ & $3.65^{+0.38}_{-0.31}$ \\ 
 $\log_{10}(L_0 / \mathrm{erg\,s^{-1}})$ & ${45.60}_{-1.40}^{+1.26}$ & ${47.67}_{-2.16}^{+1.98}$ &       $\alpha_{E_\mathrm{p}}$ & ${1.37}_{-0.75}^{+0.85}$  & ${1.44}_{-0.79}^{+0.83}$ & $1.5^{+1.0}_{-0.9}$ \\ 
 $\log_{10}(E_\mathrm{p}^\star /\mathrm{keV})$ & ${3.59}_{-0.23}^{+0.38}$ & ${3.58}_{-0.23}^{+0.26}$ &       $\beta_{E_\mathrm{p}}$ & ${0.74}_{-3.39}^{+4.53}$  & ${0.71}_{-3.37}^{+4.70}$ & $1.5^{+4.0}_{-4.1}$ \\ 
 & & &        $\theta_\mathrm{c}/\mathrm{deg}$ & ${2.87}_{-1.59}^{+2.23}$ & ${3.01}_{-1.39}^{+2.22}$ & $3.0^{+2.4}_{-2.1}$ \\
 & & &        $\theta_\mathrm{w}/\mathrm{deg}$ & ${61.39}_{-38.05}^{+27.35}$ & ${62.82}_{-40.53}^{+23.90}$ & $64.5^{+23.2}_{-42.5}$ \\  
 $\log_{10}(\sigma_\mathrm{c})$ & ${-0.07}_{-0.11}^{+0.09}$ & ${-0.07}_{-0.11}^{+0.09}$ &         $\log_{10}(\sigma_\mathrm{c})$ & ${-0.06}_{-0.08}^{+0.08}$ & ${-0.06}_{-0.09}^{+0.08}$ & $-0.40^{+0.10}_{-0.12}$ \\  
 $y$ & ${0.21}_{-0.18}^{+0.13}$ & ${0.21}_{-0.16}^{+0.15}$ &         $y$ & ${0.06}_{-0.29}^{+0.32}$ & ${0.04}_{-0.30}^{+0.30}$ & $0.0^{+0.3}_{-0.3}$ \\
 $\tau^\mathrm{min}_\mathrm{d} /\mathrm{Gyr}$  & ${0.03}_{-0.02}^{+0.28}$ & &         $\tau^\mathrm{min}_\mathrm{d} / \mathrm{Gyr}$ & ${0.04}_{-0.03}^{+0.33}$ & \\  
 $\alpha_\tau$ & ${3.20}_{-1.55}^{+1.60}$ & &         $\alpha_\tau$ & ${3.33}_{-1.59}^{+1.48}$ & \\  
 $\mu_\tau /\mathrm{Gyr}$ & & ${0.08}_{-0.07}^{+0.55}$ &         $\mu_\tau / \mathrm{Gyr}$ & & ${0.09}_{-0.07}^{+0.57}$ \\
 $\sigma_\tau$ & & ${0.11}_{-0.10}^{+1.71}$ &         $\sigma_\tau$ & & ${0.14}_{-0.12}^{+1.86}$ \\ 
 $\log_{10}(R_0 /\mathrm{Gpc}^{-3} \mathrm{yr^{-1}})$ & ${1.92}_{-0.92}^{+1.50}$ & ${2.60}_{-1.49}^{+1.23}$ &         $\log_{10}(R_0 /\mathrm{Gpc}^{-3} \mathrm{yr^{-1}})$ & ${2.57}_{-0.63}^{+0.67}$ & ${2.25}_{-0.50}^{+0.53}$ & $2.26^{+0.67}_{-0.71}$ \\
 \hline
\end{tabular}
\normalsize
\tablefoot{The \citetalias{grbpop} results refer to their `flux-limited sample' analysis.
All constraints are represented by their medians and $90\%$ posterior credible intervals.}
    \label{tab:inferred_parameters}
\end{table*}

\begin{figure*}
     \includegraphics[width=\textwidth]{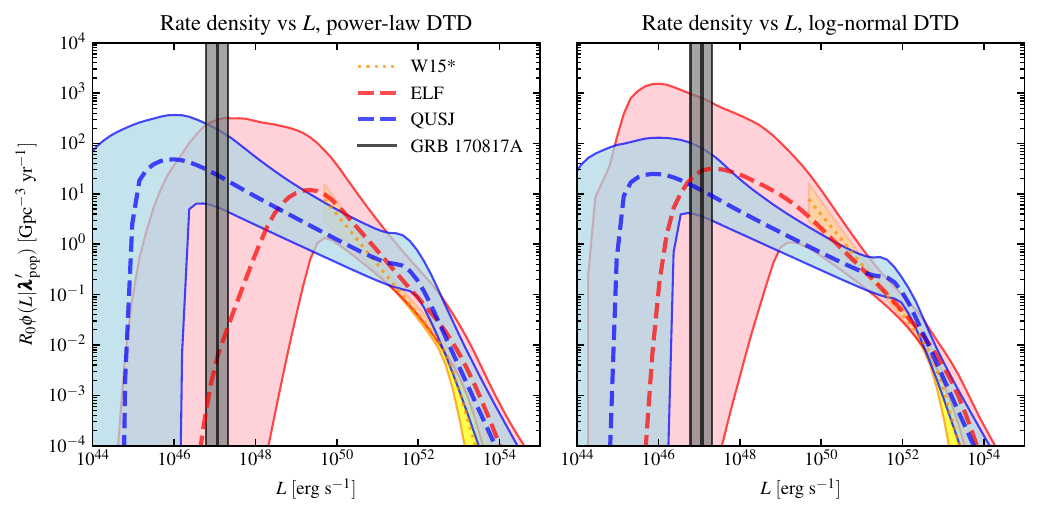}
	\caption{Luminosity probability distributions obtained considering either a power-law or a log-normal DTD (left and right panel, respectively). Results obtained with the broken power-law model are shown in red, and the ones obtained with the structured jet model are in blue. The shaded areas represent the $90\%$ credible intervals. Luminosity functions from \citetalias{WP15} are also shown in yellow. The measured luminosity for GRB~170817A is shown in grey as a reference point.}
	\label{fig:luminosity_function}
\end{figure*}

The red (resp.\ blue) filled areas in Figure~\ref{fig:luminosity_function} show the 90\% credible bands of the luminosity function obtained assuming the ELF (resp.\ QUSJ) model for the power-law DTD (left-hand panel) and the log-normal DTD (right-hand panel). In each panel, we also show the corresponding result of the \citetalias{WP15}* analysis. The luminosity of GRB~170817A \citepalias{grbpop} is shown with a vertical grey band for comparison. 

The two different DTD models considered have a limited impact on the shape of the luminosity function.  Above $L_\mathrm{iso} \sim 5 \times 10^{52}~\mathrm{erg}~\mathrm{s}^{-1}$, both of our luminosity function models show good agreement with each other and with the~\citetalias{WP15}* results. Between $L_\mathrm{iso} \sim 10^{49}~\mathrm{erg}~\mathrm{s}^{-1}$ and $L_\mathrm{iso} \sim 10^{52}~\mathrm{erg}~\mathrm{s}^{-1}$, the trend of the ELF tends to follow the one from~\citetalias{WP15}*, while the structured jet model curves have a slightly shallower slope. This is a result of the choice of parametrization and a consequence of the larger amount of information on GRB~170817A included in the QUSJ model. Below $L_\mathrm{iso} \sim 10^{49}~\mathrm{erg}~\mathrm{s}^{-1}$, the ELF model shows very large rate uncertainties compared to the structured jet scenario. This is again a consequence of the information from GRB~170817A included in that model, and in part it also reflects the stronger predictivity of the QUSJ model at low luminosity, where the minimum luminosity is set by the maximum possible viewing angle and by the slope of the apparent structure. 

\subsection{Local rate density}

\begin{figure}
	\begin{center}
        \includegraphics[width=\columnwidth]{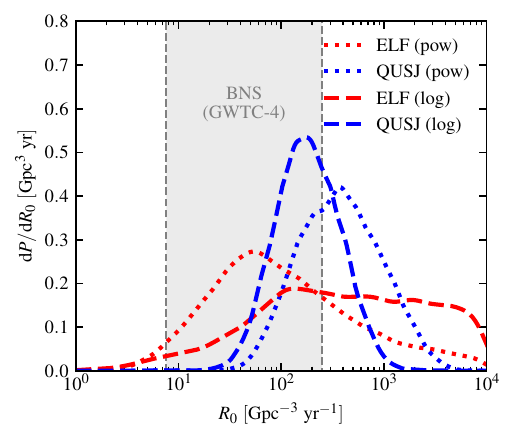}
	\end{center}
	\caption{Comparisons between local rate density probability distributions. Distributions obtained with the broken power-law and the structured jet luminosity models are shown in red and blue, respectively, while results corresponding to power-law and log-normal DTD are respectively displayed with dashed and dotted lines. The BNS rate density inferred from gravitational wave observations~\citep{GWTC4_pop} is shown in grey ($90\%$ credible intervals).}
	\label{fig:rate_comparison}
\end{figure}

As shown in Fig.~\ref{fig:rate_comparison}, the overall local rate density inferred using the ELF has a larger uncertainty compared to the result using the QUSJ model, regardless of the chosen DTD model. This is a direct consequence of the reduced predictivity of the ELF model on the minimum SGRB luminosity compared to the QUSJ model. It is instructive to compare these rate density constraints with the most recent local BNS merger rate density constraint inferred from gravitational wave observations, based on the fourth gravitational wave transient catalogue (GWTC-4; \citealt{GWTC4_pop}, $R_0 \in [7.6,~250]~\mathrm{Gpc}^{-3}~\mathrm{yr}^{-1}$ as shown by the vertical grey band in the figure). For the ELF model with the power-law DTD and the QUSJ model with the log-normal DTD, most of the posterior support is within the GWTC-4 BNS rate density constraint. For the QUSJ model with the power-law DTD, the support is concentrated at somewhat higher local rate densities, but a substantial fraction of the probability is still within the gravitational wave constraint. The ELF model with the power-law DTD rate posterior peaks within the GWTC-4 BNS rate density boundaries, reaching almost a plateau up to the upper prior PDF bound. This is a result of the poor constraints on the $L_{**}$ parameter in this case, which is correlated with $\log R_0$ (see Fig.~\ref{fig:corner_ELF_LOG} in Appendix \ref{app:corners}). In general, these results support a scenario where BNS mergers constitute the progenitors of the vast majority of SGRBs, in line with the conclusions of \citetalias{grbpop}.

\section{Discussion}~\label{sec:discussion}
The estimated time delays based on all four combinations of the models considered range between $\sim 11-18~\mathrm{Myr}$ and $\sim 764-854~\mathrm{Myr}$. The lower end of these distributions can be explained by fast BNS merger channels, which may happen in binary star systems forming with a short orbital separation and evolving quickly through a common envelope and a mass transfer phase~\citep{Tauris_2017_BNS}, where the merger time can further decrease thanks to neutron star natal kicks~\citep{Andrews_2019_kicks, Beniamini_2024_ultrafastkicks}. The parameter ranges found for the power-law DTD are compatible with those estimated from the Milky Way r-process element abundances~\citep{Chen_2025_galacticrprocessabundancy}, suggesting that BNS mergers are responsible for a significant fraction of those elements in our Galaxy.

Nonetheless, our findings for both DTD models contrast with time delay estimations in other short GRB population studies, such as~\citetalias{WP15},~\citet{Luo_2022_lognormalong} and~\citet{Tan_Yu_structured_jet}. The differences could in part be traced back to the different SGRB samples with a measured redshift considered. For example, the largest measured redshift for a GRB of the \citetalias{WP15} sample is $z = 1.131$, while our sample contains four GRBs with a measured redshift beyond that value, up to $z = 2.28 \pm 0.14$.

However, the different treatment of the selection effects could also play a role in the estimate of the SGRB time delays. On one hand, the chosen peak-flux thresholds for \textit{Fermi}/GBM in~\citetalias{WP15} and~\citet{Tan_Yu_structured_jet} and for \textit{Swift}/BAT in~\citetalias{WP15} are below the respective completeness flux thresholds. Moreover, many previous studies relied on rest-frame samples (i.e.\ samples of GRBs with measured redshifts) constructed without a careful assessment of their selection effects and how these could have distorted their redshift distribution. 
To test for the impact of incorrectly modelling the selection effects on inferring the time delay parameters, we performed the additional analysis runs described below.

\begin{figure}
\includegraphics[width=\columnwidth]{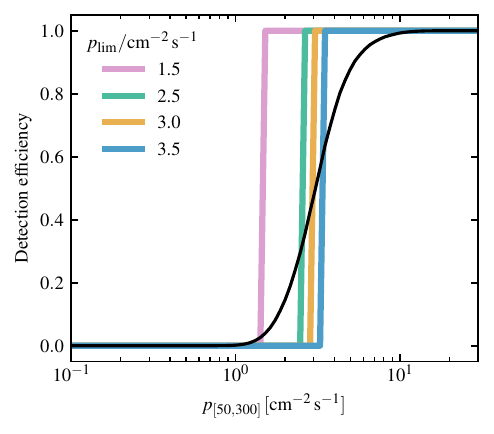}
 \caption{Comparison between the detection efficiency in our simulated sample and that used in the inference. The solid black line shows the detection efficiency model $p_\mathrm{det,GBM}$ from \citetalias{grbpop} as a function of the SGRB photon flux, assuming $E_\mathrm{p,obs}=100\,\mathrm{keV}$. The thick coloured lines show the hard-threshold detection efficiency models assumed in our inference of the simulated sample described in Sect. \ref{sec:discussion}, with different colours indicating different assumed photon flux thresholds, as in Figure \ref{fig:mockdata_dpdlogtd}. For each of these analyses, only the SGRBs above the assumed threshold were included in the inference.}
 \label{fig:pdet_plot}
\end{figure}

\begin{figure}
\includegraphics[width=\columnwidth]{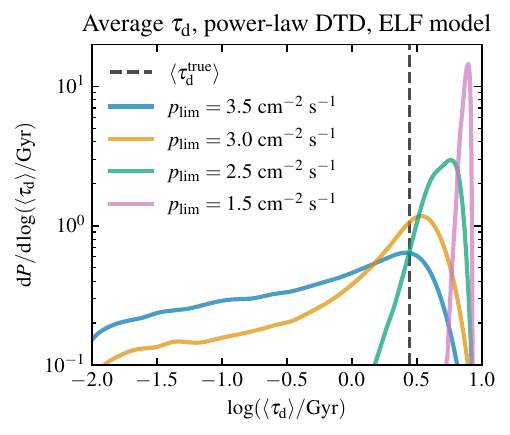}
 \caption{Posterior distribution of the average time delays inferred from a simulated SGRB population while assuming different photon peak flux cuts. The dashed black line shows the `true' average time delay value corresponding to the power-law DTD parameters from which the SGRB events have been sampled. The blue curve shows the result obtained with the same flux threshold cut as in our analysis, which ensures flux completeness of the sample. The yellow, green, and pink curves were obtained with lower threshold values, as reported in the legend.}
 \label{fig:mockdata_dpdlogtd}
\end{figure}

To demonstrate the biases deriving from using a flux-incomplete sample, we simulated a sample of short GRBs with an ELF and a power-law DTD using the following  parameters: $\alpha_\mathrm{BPL} = 2$, $\beta_\mathrm{BPL} = 3$, $\gamma_\mathrm{BPL} = 1$, $L_0 = 10^{46}~\mathrm{erg}~\mathrm{s}^{-1}$, $L_{**} = 10^{48}~\mathrm{erg}~\mathrm{s}^{-1}$, $L_* = 10^{52}~\mathrm{erg}~\mathrm{s}^{-1}$, $\tau_\mathrm{d}^\mathrm{min} = 0.1~\mathrm{Myr}$, $\alpha_\tau = 1$, $E_\mathrm{p}^\star=1\,\mathrm{MeV}$, $y=0.3$, and $\sigma_\mathrm
{c}=1$. We considered \textit{Fermi}/GBM as the only detector, and we modelled its detection efficiency following \citetalias{grbpop}. Figure \ref{fig:pdet_plot} shows the assumed detection efficiency as a function of $p_{[50,300]}$ (black line; in the plot we assume $E_\mathrm{p,obs}=100\,\mathrm{keV}$, but the dependence on the latter parameter is weak; see \citetalias{grbpop}). This allowed us to obtain a sample of 271 simulated SGRBs that were `detected'. We divided this sample into an observer-frame and a rest-frame sample of 241 and 30 events, respectively. For the inference runs, we considered a hard flux threshold detection efficiency model $P_{\mathrm{det}}(L, E_\mathrm{p},z) = \Theta\left(p_{[50, 300]}(L,E_\mathrm{p},z) - p_\mathrm{lim}\right)$ for different values of $p_\mathrm{lim}$, namely $3.5$, $3.0$, $2.5$, and $1.5~\mathrm{cm}^{-2}~\mathrm{s}^{-1}$ (the assumed detection efficiency models are shown by the thick coloured lines in Fig.\ \ref{fig:pdet_plot}). For each $p_\mathrm{lim}$ value considered, we ran the inference method only on the sub-sample of simulated SGRBs brighter than that threshold. 

Figure~\ref{fig:mockdata_dpdlogtd} shows the posterior distribution of the average time delay for each of these four inference runs. When $p_\mathrm{lim}=p_\mathrm{lim,GBM} = 3.5~\mathrm{cm}^{-2}~\mathrm{s}^{-1}$, the sample is `complete' in flux (in practice, the detection efficiency is $\gtrsim 80\%$ above the chosen threshold), and the posterior distribution peak coincides with the true value of the simulated population $\langle \tau^\mathrm{true}_\mathrm{d} \rangle = 2.78~\mathrm{Gyr}$. When the assumed $p_\mathrm{lim}$ is less than the value of $p_\mathrm{lim,GBM}$, the inference method systematically overestimates the inferred time delays. This simple simulation shows that the long time delays inferred in previous SGRB population studies might be, at least in part, the result of a bias due to an incorrect modelling of the selection effects.

\section{Conclusions}

We have performed a hierarchical Bayesian study of the short GRB population in order to determine the time delays between those events and the CSFH. To minimise observational biases, we selected both a flux-complete and a high redshift-completeness sample of SGRBs. We considered two different models for the delay time distribution, namely, a power-law with a minimum time delay and a log-normal distribution, and we modelled the luminosity probability distribution of our population either as a doubly broken power law or starting from a quasi-universal jet structure. We chose the CSFH from~\citet{MadauFragos2017}.

The inferred parameters for a power-law delay time distribution model are compatible with the estimates from the r-process elements in our Galaxy~\citep{Chen_2025_galacticrprocessabundancy}, and they correspond to time delays ranging from $\sim 12~\mathrm{Myr}$ to $\sim 686-770~\mathrm{Myr}$. The steep power-index ($\alpha_\tau > 1.5$) deviates from the `conventional' value $\alpha_\tau = 1$ adopted for a binary system that evolves while losing energy solely through gravitational radiation. This suggests that compact binary systems can merge through faster channels, for example forming with a low initial orbital separation and undergoing through a common envelope and mass transfer phase~\citep{Tauris_2017_BNS}. In some cases, the time required by the binary to merge might further reduce when neutron star natal kicks occur~\citep{Andrews_2019_kicks, Beniamini_2024_ultrafastkicks}. The same fast merger channel scenarios are compatible with our findings for a log-normal delay time distribution, where the expected time delays go from $\sim 13-14~\mathrm{Myr}$ to $\sim 877-969~\mathrm{Myr}$.

For each of the delay time distributions considered, we found that the choice of luminosity function model has a limited impact on the time delay results, supporting the conclusion that our finding does not depend specifically on our parametrization of the luminosity function.

Finally, we tested the impact of ill-modelled selection effects on the inferred time delays of a SGRB population by repeating the study for a simulated sample and using different values for its flux cut. We found that the inferred time delays become shifted towards higher values when using a flux threshold lower than the completeness one. We therefore stress the importance of building samples whose selection effects are understood and correctly modelled when performing statistical inference for parameter estimation, in order to minimise the observational biases when studying a population of astrophysical events.

\begin{acknowledgements}
    MP acknowledges the fundings from the Fonds de la Recherche Scientifique - FNRS, Belgium, under grant No. 4.4501. OS acknowledges support from the Italian National Institute for Astrophysics (INAF) through `Finanziamento per la ricerca fondamentale' grant number 1.05.23.04.04, and also funding by the European Union-Next Generation EU, PRIN 2022 RFF M4C21.1 (202298J7KT - PEACE).
\end{acknowledgements}

\bibliographystyle{aa}
\bibliography{bibliography}

\begin{appendix}

\section{Likelihoods and results for W15*}\label{app:WP15_remastered}

\begin{figure*}
	\begin{center}
        \includegraphics[width=\linewidth]{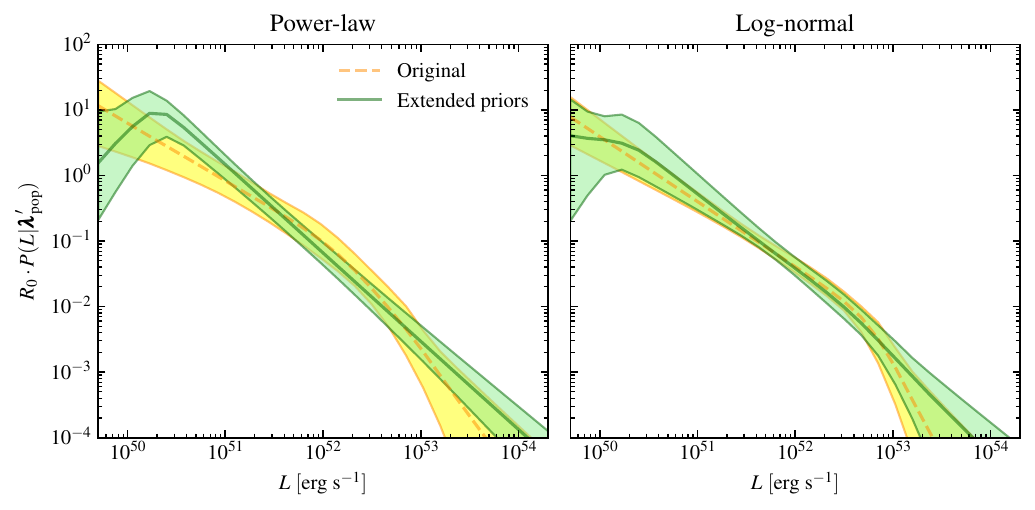}
	\end{center}
	\caption{Luminosity functions obtained with the sample and model from \citetalias{WP15}, obtained considering a power-law and a log-normal DTD (respectively left and right panel). The yellow dashed curves correspond to the original parameter range considered for the study, while the green continuous curves represent the results obtained with a wider parameter range. Shaded areas depict the $90\%$ percentiles.}
	\label{fig:wp15_lumfun_vs}
\end{figure*}

\subsection{Likelihood evaluation}\label{subsec:wp15_samples}
The likelihood numerator described in Eq.~\ref{eqn:ni_likelihood} for the \textit{Fermi}/GBM and \textit{CGRO}/BATSE samples are computed in a similar way to what done in~\citetalias{grbpop} but considering the photon spectral peak energy in the source frame fixed to $E^*_\mathrm{p} = 800~\mathrm{keV}$. We can, in fact, write the posterior probability on the source parameters such as $P(\vec{\lambda}_\mathrm{src} | \vec{d}_i) = P(\vec{d}_i | \vec{\lambda}_\mathrm{src}) \pi(L)\pi(z)/P(\vec{d}_i)$. Then, assuming that $P(\vec{\lambda}_\mathrm{src}|\vec{d}_i) \propto \pi(z)$ and neglecting the uncertainty on the measured $p_{[E_0, E_1]}$, we can write
\begin{equation}\label{eqn:p_di|lambdasrc_wp15_obsframe}
    P(\vec{d}_i|\vec{\lambda}_\mathrm{src}) \propto \frac{\delta(L-\hat{L}_i(z))}{\pi(L)} P(\vec{d}_i) ,
\end{equation}
where $\hat{L}_i(z) = L(p_{i},E_\mathrm{p}^*, z)$. We can compute the prior on $L$ by assuming a uniform prior in $p_{[E_0, E_1]}$ and applying a coordinate transform
\begin{equation}\label{eqn:wp15_coordtransf_p64_liso}
    \pi(L) = \frac{\partial p_i}{\partial L}\pi(p_i) \propto \frac{p_i}{L}    
\end{equation}
so that the numerator for the likelihood terms in the product for each event of the sample we obtain is
\begin{equation}\label{eqn:numerator_obsframe_wp15}
    \mathcal{N}^\mathrm{BATSE-GBM}_i (\vec{d}_i|\vec{\lambda}^\prime_\mathrm{pop}) = P(\vec{d}_i) \int_0^{\infty} \frac{\hat{L}_i(z)}{p_{i}} P_\mathrm{pop}(\hat{L}_i(z), z | \vec{\lambda}^\prime_\mathrm{pop}) \mathrm{d}z~.
\end{equation}

The likelihood numerator for the \textit{Swift}/BAT sample is built in an analogous way. By neglecting the uncertainty on the measured luminosity and redshift, we have
\begin{equation}\label{eqn:p_di|lambdasrc_wp15_restframe}
    P(\vec{d}_i|\vec{\lambda}_\mathrm{src}) \propto \frac{\delta(L-L_i)\delta(z-z_i)}{\pi(L, z)} P(\vec{d}_i). 
\end{equation}
If we assume a prior uniform in $\ln(L)$ and $z$, we have that $\pi(L, z) \propto L^{-1}$, and therefore the likelihood numerator for \textit{Swift}/BAT events is
\begin{equation}\label{eqn:numerator_restframe_wp15}
    \mathcal{N}^\mathrm{BAT}_i (\vec{d}_i|\vec{\lambda}^\prime_\mathrm{pop}) = P(\vec{d}_i) L_i P_\mathrm{pop}(L_i, z_i | \vec{\lambda}^\prime_\mathrm{pop}).
\end{equation}

For the events in the three samples, the likelihood denominator can be computed as in Eq.~\ref{eqn:di_likelihood}:
\begin{equation}\label{eqn:denominator_wp15}
    \mathcal{D} (\vec{\lambda}^\prime_\mathrm{pop}) = \int \Theta(p_{[E_0, E_1]}(\vec{\lambda}_\mathrm{src},E^*_\mathrm{p}) - p_{\mathrm{lim},k}) P_\mathrm{pop}(\vec{\lambda}_\mathrm{src}|\vec{\lambda}^\prime_\mathrm{pop})\mathrm{d}\vec{\lambda}_\mathrm{src}~.
\end{equation}

The astrophysical SGRB local rate density $R_0$ is estimated using the rate of observed SGRBs in the \textit{Fermi}/GBM sample, with its corresponding $\mathcal{D} (\vec{\lambda}^\prime_\mathrm{pop})$ and considering an effective observing time of $T \cdot \eta_\mathrm{DC} = 3.65~\mathrm{yr}$~\citepalias{WP15}.

\subsection{Priors and posterior evaluation}\label{subsec:wp15_priors}

The prior PDFs for the population parameters $\vec{\lambda}^\prime_\mathrm{pop}$ have been chosen independent between each other, i.e. $\pi(\vec{\lambda}^\prime_\mathrm{pop}) = \pi(\alpha_\mathrm{BPL})\pi(\beta_\mathrm{BPL})\pi(L_*)\pi(\alpha_t)\pi(R_0)$ and $\pi(\vec{\lambda}^\prime_\mathrm{pop}) = \pi(\alpha_\mathrm{BPL})\pi(\beta_\mathrm{BPL})\pi(L_*)\pi(\mu_t)\pi(\sigma_t)\pi(R_0)$ respectively for the power-law and the log-normal DTD models. All the priors have been chosen either uniform or uniform in logarithm, within the bounds that we can deduce from \citetalias{WP15} plots. The set of priors used is listed in table~\ref{tab:wp15_priors}.

\begin{table}
    \centering
    \caption{Population parameters for the \citetalias{WP15} study and priors used in our test.}
\begin{tabular}{ c l }
  Parameter & Prior \\ 
 \hline
 $\alpha_\mathrm{BPL}$ & Uniform, $\alpha_\mathrm{BPL} \in [0,~5]$ \\
 $\beta_\mathrm{BPL}$ & Uniform, $\beta_\mathrm{BPL} \in [0,~6]$ \\
 $L_*$ & Uniform-in-log, $L_* \in [10^{51},~10^{53}]~\mathrm{erg}~\mathrm{s}^{-1}$ \\  
 $\alpha_t$ & Uniform, $\alpha_t \in [0,~3]$ \\
 $\mu_t$ & Uniform-in-log, $\mu_t \in [0.01,~5]~\mathrm{Gyr}$ \\
 $\sigma_t$ & Uniform-in-log, $\sigma_t \in [0.01,~5]$ \\
 $R_0$ & Uniform-in-log, $R_0 \in [1,10^4]~\mathrm{Gpc}^{-3}~\mathrm{s}^{-1}$ \\
 \hline
\end{tabular}
    \label{tab:wp15_priors}
\end{table}

\begin{table*}
    \centering
    \caption{Constraints on parameters for the \citetalias{WP15} sample and model. Error bars depict $1\sigma$ posterior credible intervals as in \citetalias{WP15} to make a direct comparison between the parameters inferred.}
\begin{tabular}{ l | c c c c | c c }
  & \multicolumn{4}{c|}{\citetalias{WP15} prior} & \multicolumn{2}{c}{Wider prior} \\ 
  Parameter & \multicolumn{2}{c}{Power-law} & \multicolumn{2}{c|}{Log-normal} & Power-law & Log-normal \\ 
 \hline
   & This work & \citetalias{WP15} (SFR2) & This work & \citetalias{WP15} (SFR2) & \multicolumn{2}{c}{This work} \\ 
 \hline
 $\alpha_\mathrm{BPL}$ & ${0.89}_{-0.23}^{+0.15}$ & $0.90^{+0.12}_{-0.17}$ & ${0.99}_{-0.12}^{+0.09}$ & $0.96^{+0.11}_{-0.12}$ & ${-1.63}_{-0.92}^{+1.07}$ & ${0.69}_{-2.70}^{+0.36}$ \\
 $\beta_\mathrm{BPL}$ & ${2.00}_{-0.46}^{+1.00}$ & $2.10^{+1.10}_{-0.70}$ & ${2.69}_{-0.85}^{+1.55}$ & $1.90^{+1.00}_{-0.70}$ & ${1.34}_{-0.08}^{+0.10}$ & ${1.43}_{-0.19}^{+1.45}$ \\
 $\log_{10}(L_* / \mathrm{erg}\,\mathrm{s}^{-1})$ & ${52.21}_{-0.40}^{+0.55}$ & $52.30^{+0.24}_{-0.12}$ & ${52.68}_{-0.30}^{+0.22}$ & $52.30^{+0.23}_{-0.10}$ &  ${50.32}_{-0.14}^{+0.14}$ & ${1.43}_{-0.19}^{+1.45}$ \\  
 $\alpha_t$ & ${0.62}_{-0.26}^{+0.28}$ & $0.71^{+0.21}_{-0.23}$ & & & ${0.46}_{-0.28}^{+0.34}$ & \\
 $\mu_t/\mathrm{Gyr}$ & & & ${3.09}_{-0.46}^{+0.46}$ & $3.90^{+0.40}_{-0.50}$ & & ${3.02}_{-0.68}^{+0.53}$ \\
 $\sigma_t$ & & & ${0.04}_{-0.03}^{+0.12}$ & $0.00^{+0.02}$ & & ${0.07}_{-0.06}^{+0.45}$ \\
 $R_0/\mathrm{Gpc}^{-3}\,\mathrm{yr}^{-1}$ & ${13.18}_{-5.42}^{+7.23}$ & $7.7_{-4.6}^{+5.4}$ & ${7.94}_{-2.70}^{+3.28}$ & $3.6_{-1.4}^{+1.6}$ & ${16.98}_{-5.23}^{+6.46}$ & ${8.51}_{-3.02}^{+3.79}$ \\
 \hline
\end{tabular}
    \label{tab:wp15_results}
\end{table*}

As in Sect.~\ref{subsec:choice_of_priors}, the evaluation of the posterior PDF is performed through dynamical nested sampling. The number of initial live points is set to 200 times the dimensionality of the posterior PDF, for a total of $1000$ and $1200$ initial live points respectively for the power-law and the log-normal DTD models. As stopping criteria, we choose dlogz=$10^{-5}$ and 30000 effective samples and batches of live points were periodically and automatically added to reach those criteria, with each batch containing one fifth of the initial live points.

\begin{figure*}
	\begin{center}
        \includegraphics[width=\linewidth]{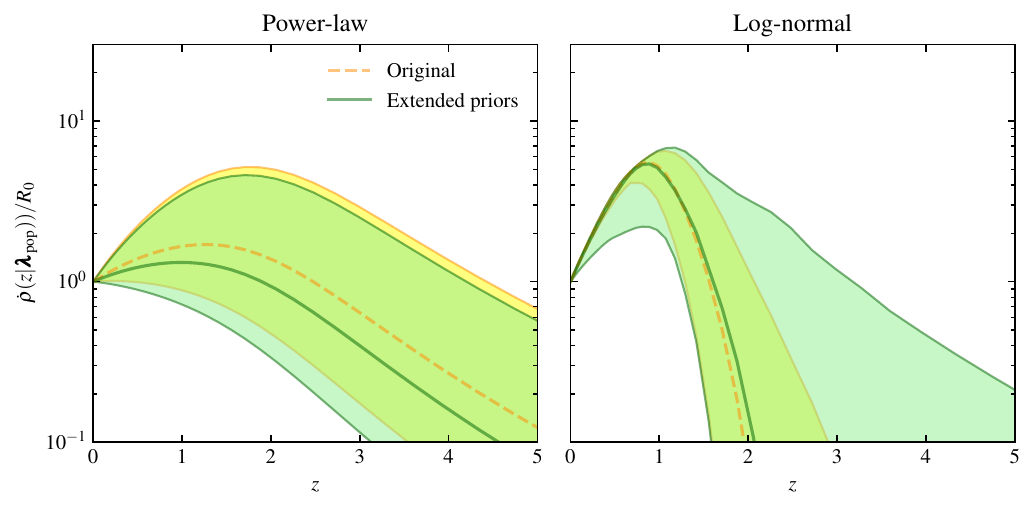}
	\end{center}
	\caption{Redshift distribution of SGRBs for the \citetalias{WP15}* population normalised to 1 at $z=0$. The power-law and log-normal DTD scenarios are shown, respectively, in the left and right panel. The yellow dashed curves depict the original parameter range considered for the study, while the green continuous curves show the results obtained with a wider parameter range. Shaded areas correspond to the $90\%$ percentiles.}
	\label{fig:wp15_rvsz_vs}
\end{figure*}

\subsection{Results}\label{subsec:wp15_results}

\begin{figure*}
    \includegraphics[width=0.49\linewidth]{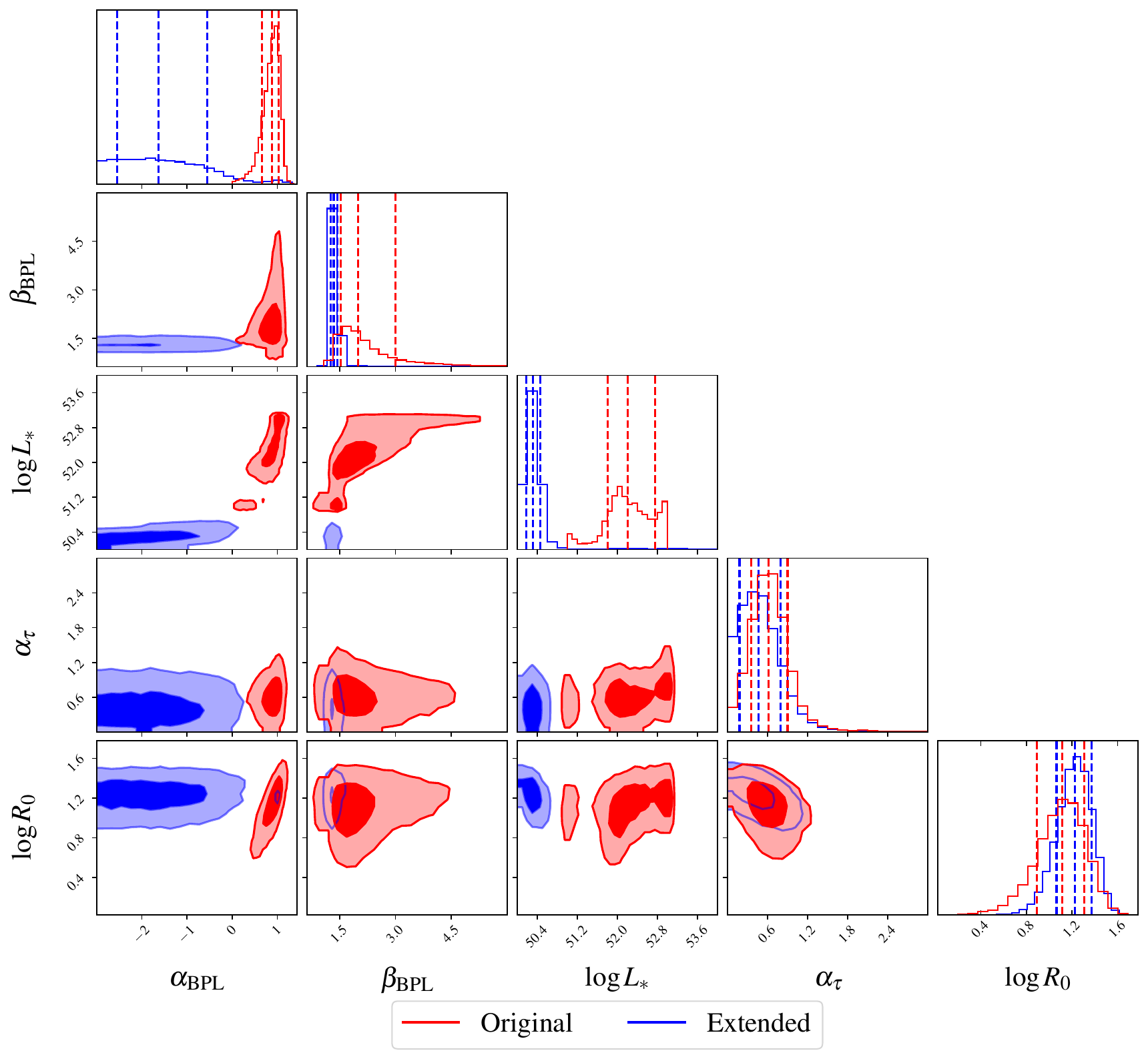}
    \includegraphics[width=0.49\linewidth]{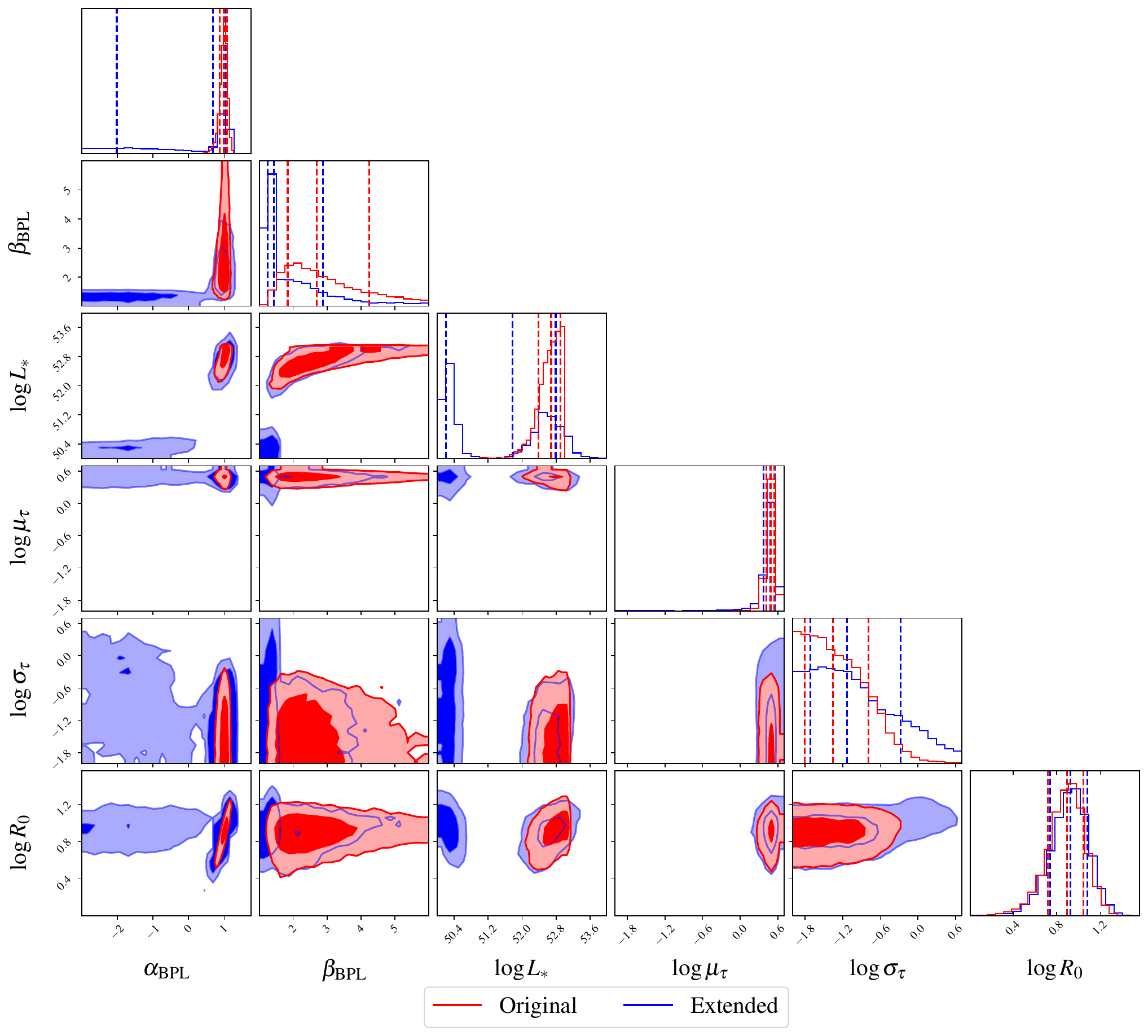}
	\caption{Posterior PDF contour plots for the \citetalias{WP15}* results along with the marginalised posterior PDFs on the sides for each parameter, in the power-law DTD (left panel) and log-normal DTD (right panel) cases. Red and blue contours respectively represent the results obtained using the "original" and extended priors on the parameters. Darker and lighter shades of the contours depict, respectively, the $50\%$ and $90\%$ confidence levels. The error bars on the marginalised posteriors mark the median of the parameters along with their $1~\sigma$ uncertainty.}
	\label{fig:corners_wp15}
\end{figure*}

The obtained posterior PDF corner plots are shown in Fig.~\ref{fig:corners_wp15} and Table~\ref{tab:wp15_results} shows the values estimated for the parameters of the population.  Although 
the CSFH we used to reproduce those results \citep{MadauFragos2017} is not featured in the \citetalias{WP15} study, its trend is very similar to the one from~\cite{PlanckSFH}, which is one of the models used for the analysis in \citetalias{WP15} and referred to as SFR2. We therefore take into account this set of results from \citetalias{WP15} to compare ours. The uncertainty taken into account on the median values is $1\sigma$ as in \citetalias{WP15}, in order to make a direct comparison between the two works.

The results are quite similar to those of \citetalias{WP15}, except for a shorter median time delay and a larger variance for the log-normal DTD and the local rate density whose estimate is one order of magnitude higher than the one in \citetalias{WP15}. We attribute the difference in the estimate of $R_0$ to the different methods used, as in \citetalias{WP15} they build their estimate on the fraction of non-collapsar events based on the criteria by~\cite{bromberg13}.

If we perform the same analysis with a wider range of parameters for our prior PDFs, we obtain considerably different results. In fact, if we consider a uniform prior in the interval $[-3,~5]$ for $\alpha_\mathrm{BPL}$ and a uniform-in-log prior for $L_{*}$ between $[10^{50},~10^{54}]~\mathrm{erg}~\mathrm{s}^{-1}$, the posterior starts to exhibit multiple local maxima as shown in Fig.~\ref{fig:corners_wp15}.

By looking at the luminosity functions obtained with these extended prior bounds (Fig.~\ref{fig:wp15_lumfun_vs}), we can see that both in the power-law and the log-normal case they peak around $L_\mathrm{iso} \sim 2 \times 10^{50}~\mathrm{erg}~\mathrm{s}^{-1}$, following a single power-law trend after.

The redshift distributions obtained, displayed in Fig.~\ref{fig:wp15_rvsz_vs}, present different behaviours for the two delay time distribution models. The power-law model with fixed minimum time delay brings to a considerable uncertainty in the peak of the distribution, ranging from $z \sim 0$ to $z \sim 2$ when considering a $90\%$ credible interval. Overall the curves look a bit shifted towards lower redshift when compared to the ones obtained with the tighter parameter range. On the other hand, the curves obtained with the log-normal model and a wider parameter range have a similar behaviour to what has been found in~\citetalias{WP15}, but with a larger uncertainty on the distribution. The larger credible intervals are related to the wider $\log \sigma_\tau$ values found with the choice of a less restrictive prior for the other parameters (see Fig.~\ref{fig:corners_wp15}).

\section{Posterior PDF corner plots}\label{app:corners}

\begin{figure*}
	\begin{center}
        \includegraphics[width=\linewidth]{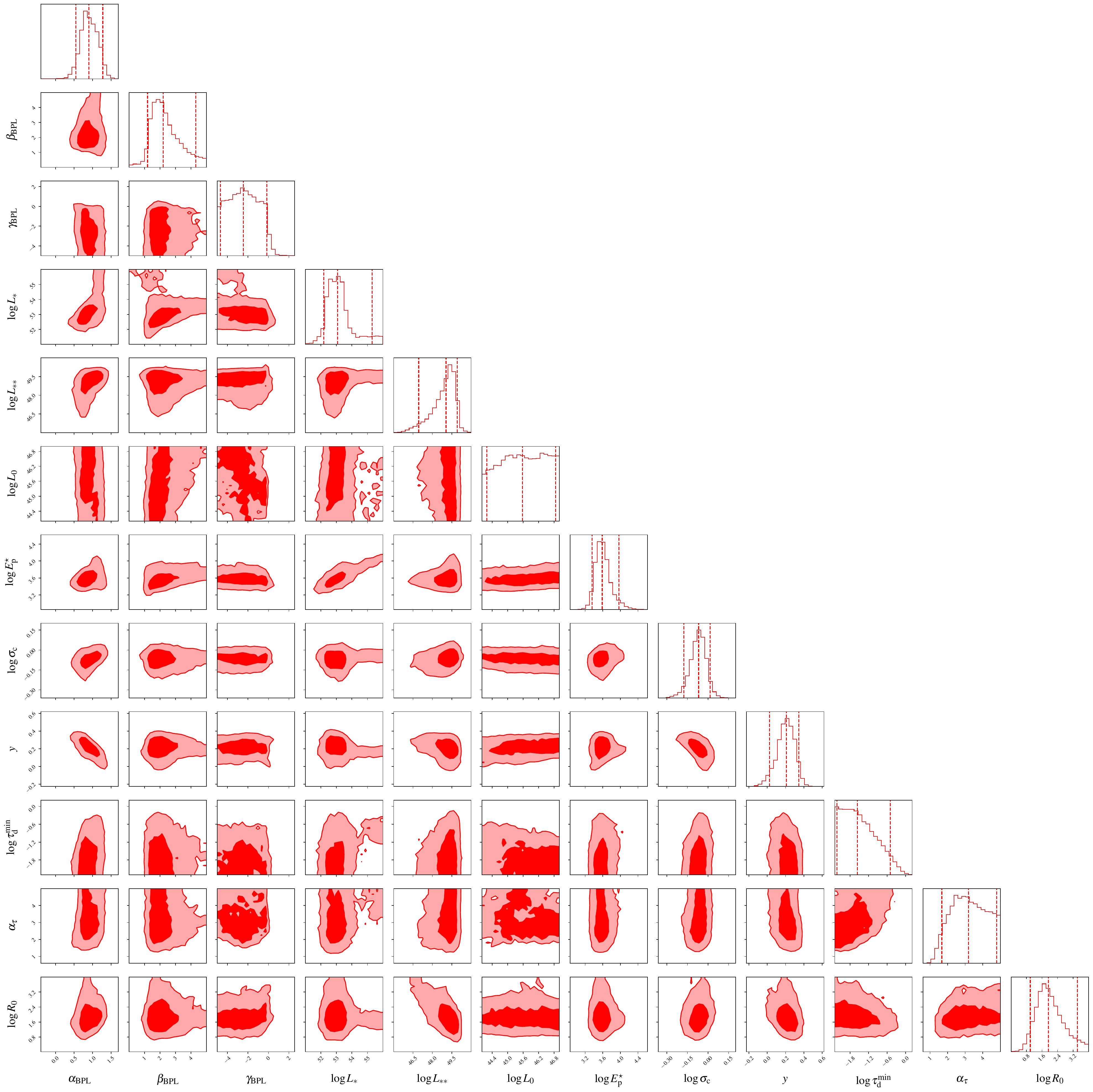}
	\end{center}
	\caption{Corner plot of the posterior PDF for the ELF + power-law DTD model, with the relative marginalised posteriors for each parameter. Contours depict the $50\%$ and $90\%$ confidence levels. Vertical dashed lines on the marginalised posteriors represent the median values with the $90\%$ credible intervals.}
    \label{fig:corner_ELF_POW}
\end{figure*}

\begin{figure*}
	\begin{center}
        \includegraphics[width=\linewidth]{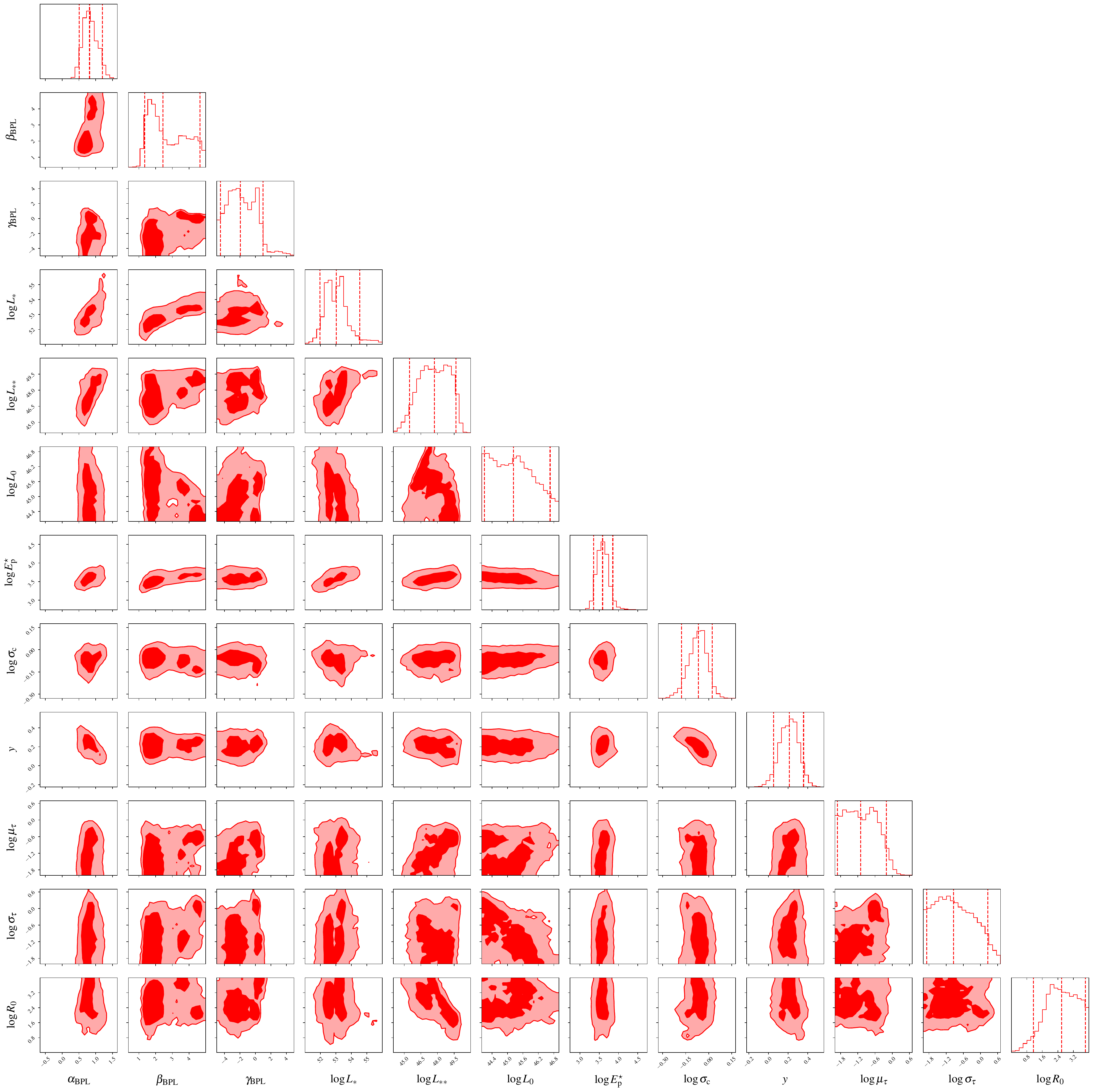}
	\end{center}
	\caption{Corner plot of the posterior PDF for the ELF + log-normal DTD model, with the relative marginalised posteriors for each parameter. Contours depict the $50\%$ and $90\%$ confidence levels. Vertical dashed lines on the marginalised posteriors represent the median values with the $90\%$ credible intervals.}
    \label{fig:corner_ELF_LOG}
\end{figure*}

\begin{figure*}
	\begin{center}
        \includegraphics[width=\linewidth]{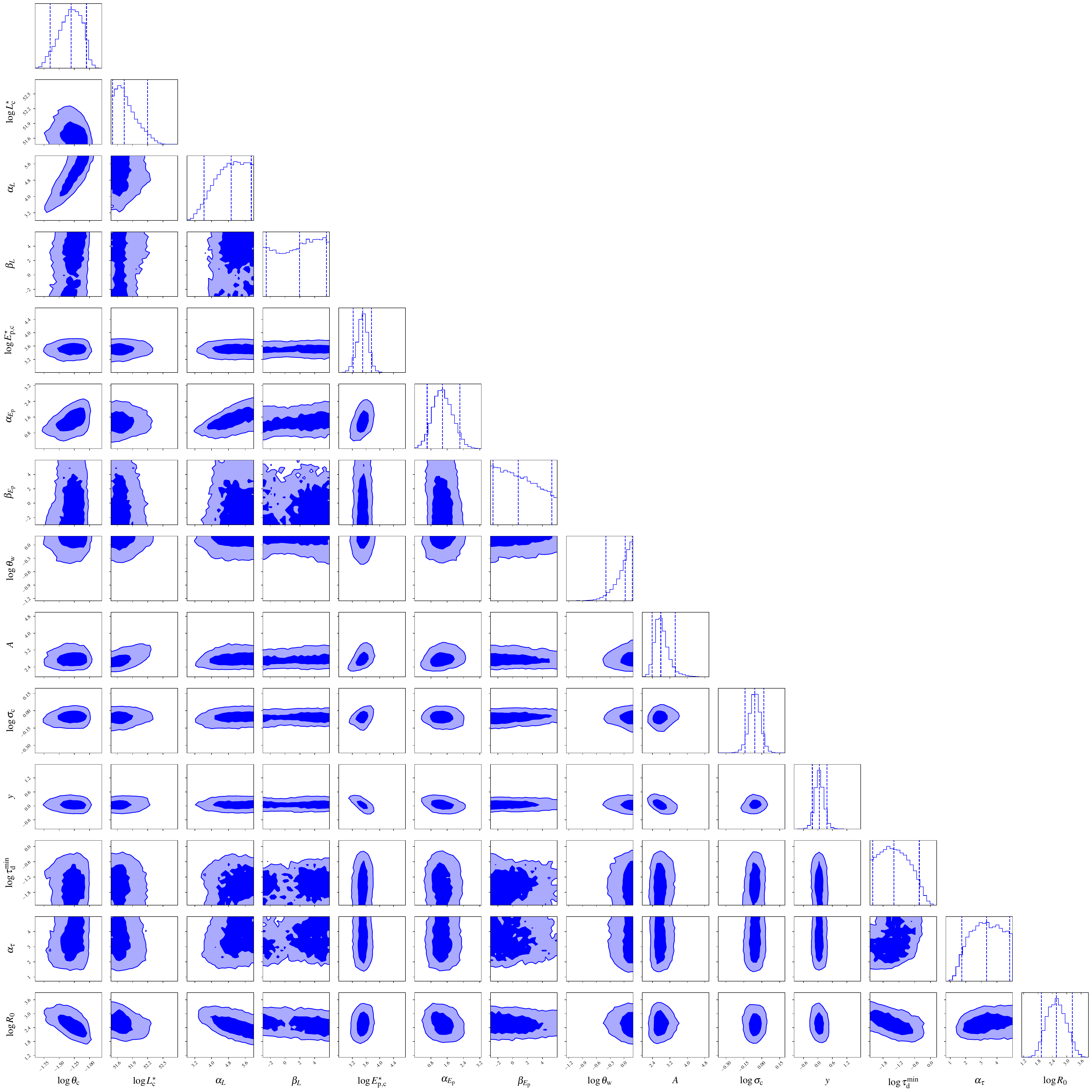}
	\end{center}
	\caption{Corner plot of the posterior PDF for the QUSJ + power-law DTD model, with the relative marginalised posteriors for each parameter. Contours depict the $50\%$ and $90\%$ confidence levels. Vertical dashed lines on the marginalised posteriors represent the median values with the $90\%$ credible intervals.}
    \label{fig:corner_QUSJ_POW}
\end{figure*}

\begin{figure*}
	\begin{center}
        \includegraphics[width=\linewidth]{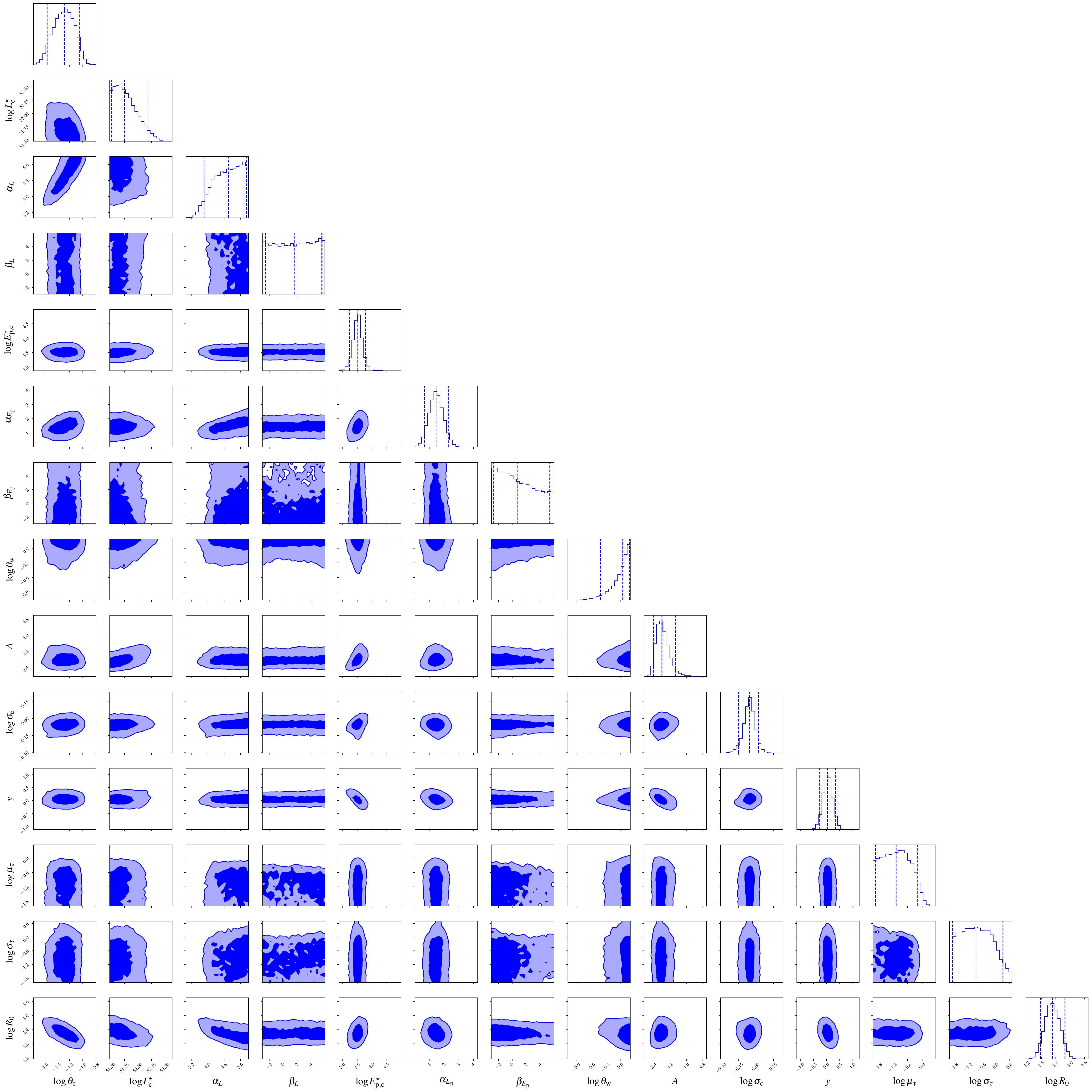}
	\end{center}
	\caption{Corner plot of the posterior PDF for the QUSJ + log-normal DTD model, with the relative marginalised posteriors for each parameter. Contours depict the $50\%$ and $90\%$ confidence levels. Vertical dashed lines on the marginalised posteriors represent the median values with the $90\%$ credible intervals.}
    \label{fig:corner_QUSJ_LOG}
\end{figure*}

\end{appendix}

\label{lastpage}
\end{document}